\documentstyle[psfig,11pt,aaspp4]{article}  % single--spaced manuscript
\def\etal{{\it et al. }}
\def\kms{km~s$^{-1}$~}

\def\W50{W$_{50}$~}
\font\eightrm=cmr8
\font\eightbf=cmbx8
\font\eightit=cmti8
\font\eightsl=cmsl8
\font\eightmus=cmmi8
\def\smalltype{\let\rm=\eightrm \let\bf=\eightbf
  \let\it=\eightit \let\sl=\eightsl \let\mus=\eightmus
  \baselineskip=9.5pt minus .75pt
  \rm}
\def \ntot{782~}
\def \nin+{584~}
\def \nin{374~}
\def \nin_use{346~}
\received{Someday 1999}
\accepted{Someday 1999}
\journalid{337}{Someday 1999}
\articleid{11}{14}

\slugcomment{To appear in {\it Astronomical Journal}}

\begin{document}
\hskip 3.5in{Version 1.1 \hskip 10pt 13 October 1996}
\title{The I--Band Tully--Fisher Relation for Cluster Galaxies: 
Data Presentation}

\author {Riccardo Giovanelli, Martha P. Haynes, Terry Herter and Nicole P. Vogt}
\affil{Center for Radiophysics and Space Research
and National Astronomy and Ionosphere Center\altaffilmark{1},
Cornell University, Ithaca, NY 14853}

\author {Gary Wegner}
\affil{Dept. of Physics and Astronomy, Dartmouth College, Hanover, 
NH 03755}

\author {John J. Salzer}
\affil {Astronomy Dept., Wesleyan University, Middletown, CT 06459}

\author {Luiz N. Da Costa}
\affil{European Southern Observatory, Karl Schwarzschild Str. 2, D--85748
Garching b. M\"unchen, Germany}

\author {Wolfram Freudling}
\affil{Space Telescope--European Coordinating Facility, Karl Schwarzschild Str. 2, D--85748
Garching b. M\"unchen, Germany}

\altaffiltext{1}{The National Astronomy and Ionosphere Center is
operated by Cornell University under a cooperative agreement with the
National Science Foundation.}

\hsize 6.5 truein
\begin{abstract}

Observational parameters which can be used for redshift--independent
distance determination using the Tully--Fisher (TF) technique are given 
for \ntot spiral galaxies in the fields of 24 clusters or groups. I band
photometry for the full sample was either obtained by us or
compiled from published literature. 
Rotational velocities are derived either from 21 cm spectra or 
optical emission line long--slit spectra, and converted to a
homogeneous scale. In addition to presenting the data, a discussion
of the various sources of error on TF parameters is introduced, and 
the criteria for the assignment of membership to each cluster are given.

\end{abstract}


\section {Introduction}

The observed radial velocity of a galaxy is 

$$cz = H_\circ d + [{\bf V}_{pec}({\bf d}) - {\bf V}_{pec}(0)]\cdot ({\bf d}/d)
\eqno (1)$$

\noindent
where {\bf V}$_{pec}$ is the peculiar velocity vector, {\bf d} is the
vector distance to the galaxy and $d$ its modulus. The Tully--Fisher (TF)
technique yields estimates of $H_\circ d$ and $cz$ can generally be obtained
directly and with high accuracy from a galaxy's spectrum. If the Cosmic
Microwave Background (CMB) radiation dipole moment is interpreted as
a Doppler shift resulting from the motion of the Local Group with respect
to the comoving reference frame, ${\bf V}_{pec}(0)$ can be inferred from
the CMB dipole and the peculiar velocity of the galaxy can be obtained
from eqn. (1), independently on any assumption regarding the value of $H_\circ$.
In this paper we shall often refer to $H_\circ d$ as ``distance'', with the 
understanding that such quantity is intended expressed in \kms.
The necessary ingredients in this operation are a set of photometric and
spectroscopic data and a template TF relation, which is empirically derived.

Clusters of galaxies are favorite targets for applications of the TF 
technique of determining redshift--independent distances $H_\circ d$ for two 
important reasons. First, a cluster provides a large number of objects located at a common distance, thereby allowing the determination of a TF relation 
slope which will be exempt from the vagaries that are introduced by an 
{\it a priori} unknown peculiar velocity field in a field galaxy sample.
Clusters are thus well suited for the generation of a TF template relation.  
Second, the combination of independent distance estimates of several 
galaxies in a cluster provides a more accurate determination of the 
cluster distance: to the extent that $N$ galaxies in a cluster can be 
considered to be at the same distance, the cluster distance can be found 
$\sim N^{1/2}$ times more accurately than as determined for a single galaxy. 
Well--sampled clusters and groups can thus provide ``hard points'' in a map 
of the large scale peculiar velocity field. 

Seminal work in the determination of redshift--independent cluster distances
was carried out in the 1980's by Aaronson and co--workers (1983,1986,1989),
using a version of the TF relation that utilized H band aperture photometry,
an approach thought to minimize the amplitude of internal extinction corrections
and therefore to reduce the scatter that determines the ultimate accuracy of
the distance determination technique. More recently, it has become common
practice to carry out surface photometry for TF work using I or R band CCD 
images. The sky background at those wavelengths is still relatively low
(as compared to H and K bands), detectors have high efficiency and large 
fields of view, and data acquisition is relatively fast even 
with small aperture telescopes. The population dominating the light at I 
band is comprised of stars that are several Gyr old. Thus, disks are well 
outlined but of smoother appearance than as seen in blue portions of the 
visible spectrum, and their apparent 
inclinations to the line of sight can be more reliably determined. In 
addition, the processes operating in clusters that may alter the star formation 
rate in galaxies will have a very retarded effect on the red and infrared 
light of disks; thus, smaller --- if any --- systematic differences are 
expected 
between the I and R TF relation of cluster and field galaxies (Pierce and 
Tully 1992). Recent CCD TF cluster work includes the surveys of Pierce and 
Tully (1988; hereafter PT88), Han (1992) Han and Mould (1992), hereafter
jointly referred to as HM, Schommer \etal (1993), 
Mathewson \etal (1992; hereafter MFB) and Bernstein \etal (1994). 

Published cluster samples suffer from a number of limitations:
 
\noindent (i) Spirals are rare in the central parts of clusters 
and, when found there, are often gas deficient, especially in the 
outer disk where rotation curves tend to flatten. As a result, HI 
and optical emission line measurements of rotational speeds are scarce and
difficult to obtain in cluster cores. 

\noindent (ii) High velocity resolution HI emission work, even
for spirals of normal gas content, becomes arduous at redshifts larger than
few thousand \kms with most radio telescopes, except for a few (Arecibo,
Nan\c cay) that are jointly unable to reach the whole sky. Thus 
galaxies with TF distance measurements tend to be the brightest in the 
cluster, and TF cluster samples mix optical and radio widths. 

\noindent (iii) In order to maximize sample size, cluster 
membership assignment has sometimes been made with relatively lax criteria:  
galaxies in TF cluster samples have tended to be representatives of an 
extended supercluster population rather than strict cluster members. 
As observations of increasing sensitivity are made and the data base of 
cluster galaxy TF distances grows, it becomes possible and desirable to apply 
stricter membership criteria, significantly improving the quality of the
determination of both the parameters of the TF template relation and of
the motions of individual clusters. 

As a complement to a program to map the local peculiar 
velocity field via TF distance measures of approximately 2000 field 
spirals (Giovanelli \etal 1994, hereinafter Paper I, Giovanelli \etal 1995: 
Paper II, Freudling \etal 1995: Paper III, Giovanelli \etal 1996a: paper IV,
da Costa \etal 1996: paper V), we present here the results of a TF 
survey restricted 
to cluster galaxies suitable both for the determination of a template 
TF relation and for the determination of the amplitude of the motions 
of the clusters themselves. Extending previous work, in this study we 
utilize a larger data base, strict membership criteria for the definition
of cluster samples, as well as recent determinations of correction recipes 
for the observed TF parameters (Papers I and II). The data base presented 
here contains both new photometric and spectroscopic data for several 
hundred galaxies, and the results of the reprocessing of relevant data  
available in the public domain. The observational data are 
utilized to produce a TF template and to investigate cluster motions 
in a companion paper (Giovanelli \etal 1996b, hereinafter referred to 
as Paper VII).

In section 2, we discuss the cluster selection and the data sets used.
In section 3, we present the individual galaxy parameters. In sections
4, 5 and 6 we discuss the procedures followed to derive disk inclinations,
velocity widths and I band fluxes. The global properties of the cluster
galaxy sample are investigated in section 7, while cluster membership 
assignment issues are discussed in section 8.

Throughout this work, we parametrize distance dependent parameters by
assuming a distance scale $H_\circ = 100 h$ \kms Mpc$^{-1}$. Whenever
explicit dependence on $h$ is not made, a value of $h=1$ is assumed.

\section {The Selection of Clusters and Data Sets}

In this section, we discuss alternative strategies for deriving the 
template relation and, in particular, the approach we have taken. Note 
that a template TF relation for the measurement of peculiar velocities 
requires an accurate estimation of a {\it velocity zero point}, which 
can be obtained independently on any assumption for, or knowledge
of the value of the Hubble constant. Ideally, the objects entering
the definition of the TF template will be globally at rest in the
comoving reference frame, assumed to be the one where the CMB dipole 
moment vanishes.

While the use of local calibrators with reliable estimates of 
primary distances is necessary if the TF relation is used for the
determination of the Hubble constant, it is of no consequence in defining 
a peculiar velocity TF template. Since only a handful of galaxies with 
primary distances exist, which can also be profitably used for TF work, 
they allow a determination of the TF zero point to no better than 
$\sim 0.1$ mag. Note that a TF template relation with $\sim 0.1$ mag 
distance modulus predictive accuracy is clearly unsatisfactory for the
measurement of the velocity field, as it would introduce large, spurious 
and systematic peculiar velocities of growing amplitude in the outer 
regions of any sampled volume.

\subsection {Single Cluster Template}

One commonly--adopted approach to the determination of
a TF template relation is to select a single cluster of galaxies 
as a reference, thereby equating the universal template with the
TF relation defined by its member galaxies (PT88;
MFB). In this instance, all
cluster members are assumed to be at approximately the same distance 
from the observer as predicted by the systemic velocity of the cluster,
and their relative locations in the TF diagram are 
assumed to be unaffected by peculiar motions. This strategy 
requires that the cluster be endowed with several special
qualities: (a) a substantial number of individual galaxy 
distances must be measurable for that cluster, either by virtue of its 
large spiral population or of its proximity; (b) there must
be a reasonable expectation that the cluster is  
at rest in the comoving reference frame; (c)  there must be strong 
reasons to believe that the cluster's TF relation is a particularly 
good approximation to the true one. 
The practical advantages of having a TF relation defined by a large sample
of objects, as required by (a), are obvious; the likelihood that a 
reliable template can be obtained from a {\it single} cluster is however 
questionable on statistical grounds. 

Consider the case of a nearby cluster first. We assume 
that a cluster TF relation is a 
random realization of some universal relation, defined by a zero 
offset and a slope. Uncertainty is contributed by the limited accuracy 
of measurements, by an intrinsic amount of scatter in galaxies' 
properties and by the {\it a priori} unknown amplitude of the cluster 
motion with respect to the comoving reference frame. If the number
of observed galaxies is large and the dynamic range in the observed 
parameters high, then the slope of the universal relation should be 
increasingly well approximated by that of the chosen cluster. These 
requirements favor the selection of a nearby cluster in which the fainter 
galaxies are more easily observable. The uncertainty on the offset or 
zero point of the TF relation can also be initially reduced by increasing
the size of the sample. However, the limit of this process depends on the 
amplitude of the cluster motion itself, and for a given amplitude,
the introduced uncertainty increases with cluster proximity.
In other words, if the chosen cluster is nearby, the TF zero point 
uncertainty remains very high, independent of the size of the sample. 
Note that the uncertainties on the TF zero point and slope are coupled 
quantities. 

Consider now the case of a distant cluster, where the average single--point
scatter about a best fit linear relation is on the order of a 
third of a magnitude. A typical TF cluster sample may include two dozen
galaxies, so that the cluster zero offset can be determined to better
than 0.1 mag. A cluster motion of 300 \kms --- about half the velocity 
of the LG with respect to the CMB reference frame --- translates into 
an offset of the cluster TF relation smaller than 0.1 mag only if the 
cluster is at $cz > 6000$ \kms. There is no {\it a priori} validity to 
expecting that the desired quality (b) listed above may hold for any given 
cluster. If we were to use a single cluster as reference, we would then be 
forced to use a distant one, which would limit both the number and the 
dynamic range in luminosity of the sampled galaxies. 

\subsection {``Basket of Clusters'' Template}

A more attractive approach is that of building a template by combining data 
of many clusters. It is reasonable to expect that the mean motion of a 
``basket of clusters'', if well distributed over all the sky, approximates 
null velocity better than any single cluster. The TF relation parameters 
can be statistically better determined, as the larger number of objects sampled 
help to reduce statistical uncertainties. Nearby clusters in the set
play an important role in the definition of the template slope, while
the more distant clusters allow a fairer definition of the kinematical
zero point. In order for the TF relation of each cluster to be adequately 
``spliced'' to that of the others, it is necessary to ascertain its peculiar 
velocity with respect to the reference frame defined by the cluster set. 

Convinced of the superiority of this approach, we have thus assembled TF data 
for 24 clusters of galaxies. In their majority, the data are new, and obtained 
by us at the observatories of Arecibo, Palomar, MDM, Kitt Peak, Cerro Tololo, 
Green Bank and Nan\c cay. The raw data, details of the observing schedules 
and data reduction techniques will be given in separate papers. Here we 
present the final parameters, in a stage ready for utilization in astrophysical 
applications.

\subsection {Cluster Selection}

The clusters are all relatively nearby; their systemic
velocities, as measured in the CMB reference frame, are all less than 10,000 
\kms. They span a large range in richness: the sample includes dynamically
evolved structures such as the Coma cluster and loose groups such as MDL59. 
It should be admitted that an important criterion for selection was one of 
expediency, i.e. a group or cluster may have been included in the sample if 
high quality velocity widths and I band photometry were available for a 
substantial number of member galaxies. We have not used certain clusters or 
groups, which have appeared in other cluster 
lists, for a variety of reasons. If photometry was not available in the I band, 
and thus it could not be reprocessed according to the same precepts followed for
our sample, we preferred to forego the uncertain task of cross--calibration
of data of various optical bands.
In few cases in which I band data was available, after our cluster membership
criteria were applied the number of remaining galaxies with measured distances 
for the cluster or group was so small (or cluster definition so uncertain) that
the structure did not constitute a worthwhile addition. For example, the 
cluster A534, for
which a significant amount of data is available, including our own, is at very 
low galactic latitudes. Although the size of the data set and cluster definition
would warrant its use, the very high uncertainty and clumpiness in the
extinction arising within our Galaxy make the data set currently unfit for TF 
calibration use. A significant amount of data is also available for
the Sculptor group. This is a nearby group, for which a Cepheid distance
modulus of 26.5$\pm0.2$ has been measured for one of its brightest members 
(NGC 300; Freedman \etal 1991); however, the group appears to be quite 
extended along the line of sight, perhaps as much as 2 mag in distance 
modulus (Pierce and Tully 1992). We have thus chosen to exclude it from 
our sample. The Virgo cluster, nearby and well studied, exhibits very
significant substructure, and will be the object of a successive focused
study by our group.

We were also guided by the second important criterion that the sample should
be as balanced as possible between contributions from different parts of the 
sky. Spatial balance in the cluster sample is required so that the kinematical 
zero point of the TF template, the derivation of which we describe in Paper 
VII, will be as closely as possible consistent with null velocity with 
respect to the CMB. We have been partly 
succesful, although the more distant clusters in the sample are principally 
located in the Northern hemisphere. Three stereographic views of the 
distribution of the 24 clusters are presented in figure 1, where their 
locations are in Cartesian Supergalactic coordinates $(X,Y,Z)$, expressed 
in \kms as measured in the CMB reference 
frame (Kogut \etal 1993). The conversion of cluster centers and systemic 
velocities to $(X,Y,Z)$ is done using the apex parameters as given in the
RC3 (de Vaucouleurs \etal 1993), and it does not correct the 
coordinates for the clusters' peculiar velocities.

Table 1 lists the main parameters of the chosen clusters. Standard names
are listed in column 1; in a few cases (NGC 383, NGC 507, NGC 3557), the
group is identified by the name of the most prominent galaxy. Adopted
coordinates of the cluster center are listed in columns 2 and 3, for the
epoch 1950; they are in general obtained from the same sources yielding
cluster redshifts, except in a few cases discussed in Section 8. The 
systemic velocities, heliocentric and within 
the CMB reference frame, are listed in cols. 4 and 5 respectively. 
An estimated error for the systemic velocity is given between brackets
after the heliocentric figure.
Sources for centers and systemic velocities are
given in col. 11. Angular and Cartesian supergalactic coordinates are
given in cols. 6 and 7, and in cols. 8--10, respectively. When a source
for the cluster parameters is not given in Table 1, it should be assumed that
we have rederived them using compilations of redshift data, which
occasionally include a richer data set than used in the most recent published
analyses of those groups or clusters. Available redshift data have also been 
used in the assignment of cluster membership to individual galaxies.

For the I band photometry, we adopt data from several sources. The main 
ones are: (a) our own survey of cluster and field objects, carried out using 
the MDM 1.3m and the Kitt Peak 0.9m telescopes (the raw data and the technical 
details of these observations will be presented elsewhere); (b) the data sets
published by HM and (c) that of MFB. In the case of the Ursa Major 
cluster, we make extensive use of the data of PT88.

For the velocity width data, the main sources of observational material are:
(a) our own 21 cm and optical rotation curve data obtained at several 
telescopes, namely Arecibo, Green Bank (both 300 foot and 140 foot telescopes), 
Nan\c cay, Effelsberg and Palomar (the raw data and the technical details of these observations will be presented elsewhere); (b) the data
set of MFB and of numerous other sources of 21 cm widths.

Observed raw parameters of all data sets have been reprocessed, converting 
them into a homogeneous sample, to which a consistent set of correction
algorithms has been applied. Details 
of those corrections are given in sections 4 to 6. 

\section {Galaxy Data}

In Table 2\footnote{Table 2 is presented in its complete form in the
AAS CD--ROM Series, volume 8, 1997. The first page of 23 is presented 
here for guidance regarding its form and content. Electronic copies 
can be obtained from the first author.}
 we list parameters of individual galaxies for each cluster
set. Data for a total of \ntot galaxies are presented. Of those, 374
are classified as bona fide cluster members, and an additional 210
are ``peripheral'' cluster members as defined below. The remainder
of 198 galaxies consists of either foreground or background field objects,
or pertains to groups or clusters not used in the TF template analysis.
Figures 2 through 12 display their spatial and velocity locations. 
For each cluster two plots are given: in the upper panel the distribution 
of objects in the sky is shown, while in the lower panel each galaxy's
radial velocity is plotted versus the angular distance from the cluster
center. For each cluster field, galaxies with measured TF distances are
separated into three main classes: 
(i) objects that, on the basis of 
positional and velocity data, can be assigned cluster membership (we refer
to these as the {\bf in} sample); they are generally identified by filled 
circles in figures 2 through 12; 
(ii) objects with velocities very close to the systemic velocity of the 
cluster but spatially removed from their center so that a firm membership 
assignment cannot be made; they are identified by unfilled circles in 
Figures 2 through 12 and will sometimes be referred to as ``peripheral'' 
cluster objects; the combination of these objects with those in the 
{\bf in} sample will be said to form the {\bf in+} sample; 
(iii) foreground and background galaxies are generally plotted using
asterisks. In some of the cluster plots the object identification
may be slightly more complicated, as for example when more than one
cluster is included in a single figure: see then the relevant figure 
caption, the description of col. 7 in Table 2 below and the comments
for individual clusters in Section 7. 
Galaxies with redshifts known to us but with no measured
TF distance (i.e. lacking I band photometry or an accurate estimate of width)
are plotted in Figures 2 through 12 as dots. In the upper
panel of each figure, circles with radius of respectively 1 and 2 Abell
radii are plotted as dashed lines. In the lower panel of each figure,
a solid horizontal line identifies the adopted systemic velocity of the
cluster. In a few cases of clusters with a very rich redshift data
base, caustic lines computed for $\Omega = 0.3$ are plotted as solid
curves: for them the assignment of objects to the {\bf in} sample is 
extended farther from the center of the cluster than for other clusters.

Entries in Table 2 are sorted first by the Right Ascension of each cluster, 
and within each cluster sample by increasing galaxy Right Ascension. The 
listed parameters are:

\noindent 
Col 1: one or two identification names. If the galaxy is listed in the UGC
catalog (Nilson 1973), the UGC number is listed first; if not, the first
listing corresponds to our own internal coding number, which we maintain for
easy reference in case readers may request additional information on the object.
The second galaxy name is added with the following priority scheme: if the 
galaxy has an NGC or IC number, that coding is entered preferentially; if 
the galaxy is listed in the CGCG (Zwicky \etal 1968), that coding is 
listed with second priority, in the form: field number -- ordinal number 
within the field; for southern galaxies, if an NGC or IC number is not 
available, the Lauberts (1982) catalog name is entered. If entries in those 
four catalogs are not available, a second name is not entered.

\noindent
Cols. 2 and 3: Right Ascension and Declination in the 1950.0 epoch, as listed
by the observers or available in our compilation, whichever has higher accuracy;
typically, the listed positions have 15" accuracy, including those in Southern
clusters as reported by MFB (most of the positions were conveyed by us to MFB
ahead of their observations).

\noindent
Col. 4: morphological type code in the RC3 scheme, where code 1 corresponds 
to Sa's, code 3 to Sb's, code 5 to Sc's and so on. When the type code is 
followed by a ``B'', the galaxy disk has an identifiable bar.

\noindent
Col. 5: the galaxy radial velocity as measured in the CMB reference frame
(Kogut \etal 1993); this value is obtained from the source for the radial 
velocity and velocity width data, as specified in col. 16.

\noindent
Col. 6: the angular distance $\theta$ in degrees from the center of each 
cluster. When the cluster consists of several subclumps, such as in Cancer, 
the distance is measured from the center of the main clump (clump A in 
Cancer; see Bothun \etal 1983). This parameter is the abscissa in the
lower panel of Figures 2 through 12.

\noindent
Col. 7: membership code. In most cases the code consists of one of 4 letters:
{\it c, g, f, b}. Code {\it c} signifies that the galaxy is considered a
{\it bona fide} cluster member ({\bf in} sample). Code {\it g} means that 
the galaxy is sufficiently removed from the cluster center that cluster 
membership cannot be safely assigned; the galaxy is however in the periphery 
of the cluster, at roughly the same redshift and it is a nearby supercluster 
member ({\bf in+} sample). Codes {\it f} and {\it b} identify
respectively foreground and background galaxies; these will not be used in 
the construction of the cluster TF relation in any form. Criteria adopted  
to arrive at the membership assignment are discussed in greater detail in 
the section 8. In some clusters, additional codes had to be used, due to 
the complex nature of the structure. We describe those codes separately 
below.

In Cancer, the cluster has been recognized to be an assembly of clumps, dubbed 
A, B, C, D (Bothun \etal 1983). We use the criteria given in that reference to
assign membership to each of the clumps, and used correspondingly capital 
letters A, B, C and D as membership identifiers. In the remainder of this 
paper, when mention is made to ``Cancer cluster members'', we refer to 
members of clump A, the largest of the system. In Figure 5, members of the
other clumps in the region are plotted as unfilled squares (D clumo)
and triangles (clumps B and C).

In Centaurus (A3526), Lucey \etal (1986 and references therein) recognized 
the existence of two structures superimposed along the line of sight, 
concentric in projection, with systemic velocities respectively near 
3,000 \kms and near 4,500 \kms, thus referred to as Cen30 and Cen45. 
We assign the {\it c} membership code to galaxies deemed to be part of 
Cen30, the low velocity cluster. We don't use galaxies in Cen45 for the
construction of the TF template. We list separately objects in Cen45 
(labelled as ``4'' in the relevant section of table 2) with foreground and 
background objects. The 
distinction between Cen30 and Cen45 is uncertain, especially in regions 
close to the cluster center, where virial distortions produce mixing of the
two velocity fields. In cases of greater uncertainty, we have assigned 
membership preferentially to Cen45, in order to keep the Cen30 sample as 
``clean'' as possible. It is thus likely that some of the objects assigned 
to Cen45 are Cen30 members; they are thus footnoted. In the remainder of 
the paper, when A3526 cluster 
members are mentioned, we refer to members of the Cen30 concentration.
In Figure 8, objects assigned to Cen45 are plotted as unfilled squares.

In the A3574 (Klemola 27) field, a smaller, low velocity dispersion group 
is also present, identified by Abell, Corwin and Olowin (1989) as S0753. Three 
objects in 
the field are identified as members of that group and labelled with the letter 
{\it s}. As with background and foreground objects, no further TF use is made 
of those three galaxies. In the sky plot of A3574 in Figure 9, a rich cluster is 
clearly discernible approximately 4$^\circ$ NW of the center of A3574: 
it is A1736, at a redshift $z\sim 0.04$.

Clusters A2197 and A2199, which are quite close in the sky, are listed jointly, 
so that peripheral galaxies, i.e. those of code {\it g}, are assigned 
to the binary cluster system rather than to either one of the two; angular
distances $\theta$ for these objects are referred to the center of A2199. Members 
of A2197 are identified by a ``7'' and members of A2199 are identified by a ``9''.

The case of A2634 (near $cz_{cmb} \sim 8900$ \kms) and A2666 (near 
$cz_{cmb}\sim 7800$ \kms) is the most complicated. Scodeggio \etal (1995) 
have produced a detailed study of the region, and within the redshift range 
spanned by the two clusters they also distinguished several other groups. In 
particular, one of the groups is at a systemic velocity near 7,000 \kms, near
in redshift to, but distinct from A2666. We reserve the code ``4'' for 
the {\it bona fide} members of A2634, a 
code ``6'' for the members of A2666, and the code ``7'' for the members of 
the 7,000 \kms group projected on the foreground of A2634. The code {\it g4} 
is reserved for peripheral members of A2634 and {\it g6} for the one peripheral
member of A2666.

\noindent
Col. 8: the measured velocity width, as reported by the source (see comments
to col. 16 for details), in \kms. Radio widths refer to values measured at
a level of 50\% of the profile horns.

\noindent
Col. 9: the velocity width in \kms after all corrections except that for 
inclination are applied; these corrections, including those for instrumental 
and data processing broadening, signal to noise effects, insterstellar 
medium turbulence, shape of the rotation curve and cosmological stretch, 
are discussed in section 5.

\noindent
Col. 10:  the corrected velocity width converted to edge--on viewing, in \kms.

\noindent
Col. 11: the adopted inclination $i$ of the plane of the disk to the line of 
sight, in degrees, (90$^\circ$ corresponding to edge--on perspective); the 
derivation of $i$ and its associated uncertainty are discussed in section 4.

\noindent
Col. 12: the logarithm in base 10 of the corrected velocity width (value in 
col. 10), together with its estimated uncertainty between brackets. The 
uncertainty takes into account both measurement 
errors and uncertainties arising from the corrections. The format 2.228(11), 
for example, is equivalent to 2.228$\pm$0.011.

\noindent
Col. 13: the measured I band magnitude, extrapolated to infinity assuming 
that the surface brightness profile of the disk is well described by an 
exponential function; this value is 
sometimes adopted as an average of two or more measurements, as footnoted.

\noindent
Col 14: the apparent magnitude, to which k--term, galactic and internal
extinction corrections were applied; details on the adopted corrections are given in 
section 6. 

\noindent
Col. 15: the absolute magnitude, computed assuming that the galaxy is at the 
distance indicated either by the cluster redshift, if the galaxy is a cluster 
member
({\bf in} sample), or by the galaxy redshift if not. The calculation assumes 
$H_\circ = 100h$ \kms Mpc$^{-1}$, so the value listed in column 15 is strictly 
$M_{cor} + 5\log h$. In calculating this parameter, radial velocities are  
expressed in the CMB frame and uncorrected for any cluster peculiar motion. 
The uncertainty on the magnitude, indicated between brackets in hundredths
of a mag, is the sum in quadrature of the 
measurement errors and the estimate of the uncertainty in the corrections 
applied to the measured parameter, as described in section 6. The error
estimate does not include the uncertainty on the value of the distance.

\noindent
Col. 16: Finally, references are given for the sources of the data and  
cautionary notes for individual objects. A letter code identifies the source 
for the photometry, as follows:

$a$ refers to our own photometric observations;

$b$ refers to photometric observations of Bernstein \etal (1994), restricted 
to Coma;

$h$ refers to photometric observations reported by HM;

$m$ refers to photometric observations of MFB;

$p$ refers to photometric observations of PT88, restricted to Ursa Major;

$s$ refers to photometric observations of Bureau \etal (1996; BMS), 
restricted to Fornax;

$x$ refers to a combination of different sources, as noted.

\noindent A numerical code identifies the source for the velocity width:

0 refers to our own 21 cm measurements;

1 refers to our own optical spectroscopy;

2 refers to averages of several measurements; see notes in each case; 

3 refers to 21 cm observations of Gavazzi (1987) and Scodeggio and Gavazzi (1993); 

4 refers to 21 cm observations of Bothun \etal (1985), Aaronson \etal (1989),
cross--referenced with the compilation of Bottinelli \etal (1990);

5 refers to 21 cm measurements of MFB;

6 refers to optical spectroscopy of MFB;

7 refers to 21 cm measurements reported by PT88;

8 refers to 21 cm observations of Fontanelli (1983), Lewis (1985) and 
Schneider \etal (1991);

9 refers to 21 cm observations of Bureau \etal (1996).

When an asterisk appears at the end of the record, a detailed comment is 
given for that particular object. Because of the length and number of these
comments, they are not appended to the table but included in the text as
follows. When two parameters are given in the notes for a given source,
e.g. ``HM (13.79,78)'', the first number refers to the raw magnitude and 
second to the disk inclination.

\vskip 0.25in

\noindent {\bf N383 Group:}

{\smalltype 
\noindent 453: tiny disk; exponential disk over narrow range of radii; photometric
parms. uncertain.
 
\noindent 100561: v. uncertain ellipticity.

\noindent 557: 21cm line disturbed: comp. superimposed; galaxy not used for fits.

\noindent 556: 21cm line disturbed: comp. superimposed? galaxy not used for fits.

\noindent 565: discrepancy between HM (13.79,78) and us (13.72,75): use (13.75,75).

\noindent 575: discrepancy between HM (13.77,90) and us (13.87,84): use (13.82,84).

\noindent 629: note low i; not used in fits.

\noindent 632: marginal 21cm spectrum.

\noindent 633: discrepancy between HM (12.94,80) and us (13.02,80): use (12.98,80).

\noindent 673: discrepancy between HM (14.25,71) and us (13.86,70): use (14.06,70).

\noindent 697: marginal 21cm spectrum.

\noindent 724: note low i.

\noindent 110090: tiny disk; exponential disk over narrow range of radii; photometric
parms. uncertain.
}

\vskip 0.15in\noindent {\bf N507 Group:}

{\smalltype 
\noindent 800: note low i. 

\noindent 809: discrepancy between HM (13.59,90) and us (13.50,83): use (13.55,83).

\noindent 810: discrepancy between HM (13.11,79) and us (13.04,78): use (13.07,78).

\noindent 1013: discrepancy between HM (11.09,74) and us (11.20,72): use (11.15,73).

\noindent 110334: tiny disk; exponential disk over narrow range of radii; photometric
parms. uncertain.

\noindent 110363: tiny disk; exponential disk over narrow range of radii; photometric
parms. uncertain.

\noindent 1094: discrepancy between HM (11.91,80) and us (12.01,79): use (11.96,80).
}

\vskip 0.15in\noindent {\bf A262:}

{\smalltype 
\noindent 110463: tiny disk; exponential disk over narrow range of radii; photometric
parms. uncertain, ellipticity may be underestimated.

\noindent 1234: discrepancy between HM (13.63,51) and us (13.50,64) is large; not sure we were measuring same galaxy; use our parms.

\noindent 1307: discrepancy between HM (12.07,81) and us (12.29,79): use (12.18,80).

\noindent 1316: unclear ID of these I band data: U1316 or 522-023? assign to U1316 but caution.

\noindent 1376: i uncertain; inner parts affected by bar asymm., suggest high i; outer 
isophotes much lower i; adopt low i of outer.

\noindent 1405: wiggly photometric profile; total magnitude quite uncertain.

\noindent 1416: galaxy is 1.5 mag off mean TF relation: misident? not used in fits.

\noindent 1456: large uncertainty on i affects mostly width.

\noindent 1729: tiny disk; exponential disk over narrow range of radii; photometric 
parms. uncertain; unstable ellipticity.
}

\vskip 0.15in\noindent {\bf A400:}

{\smalltype 
\noindent 2336: severe TF outlier; not used in fits.

\noindent 2364: marginal membership assignment.

\noindent 2367: Type S0a: uncertainty in corrections large.

\noindent 2375: discrepancy between HM (12.93,76) and ours (13.17,81): use (13.05,79).

\noindent 2405: discrepancy between HM (12.97,72) and ours (13.08,70): use (13.03,70).

\noindent 2454: discrepancy between HM (13.87,85) and ours (13.89,79): use (13.88,80).
}

\vskip 0.15in\noindent {\bf Eridanus:}

{\smalltype 
\noindent 22364: note low i; not used in fits.

\noindent 22413: note low i; MFB give i=49, but their major and minor axis data yield i=42; we adopt i=45.

\noindent 22472: note low i; MFB give i=49, but their major and minor axis data yield i=39; we adopt i=44.

\noindent 22479: width v. small: low reliability.

\noindent 22625: width v. small: low reliability.

\noindent 22735: width v. small: low reliability.

\noindent 430438: discrepancy between MFB (11.67,89) and us (11.75,90): use (11.71,90)

\noindent 430459: discrepancy between MFB (11.75,75) and us (11.62,73): use (11.68,73)

\noindent 23183: width v. small: low reliability.

\noindent 23560: width v. small: low reliability.

}

\vskip 0.15in\noindent {\bf Fornax:}

{\smalltype 
\noindent 22003: note low i; unreliable width; not used in fits.

\noindent 22216: width v. small: low reliability.

\noindent 22370: width v. small: low reliability.

\noindent 22382: width very uncertain; measured on paper plot.

\noindent 22447: discrepancy between BMS (12.46,52) and MFB (12.69,46): use (12.55,50).

\noindent 22474: discrepancy between BMS (13.15,84) and MFB (13.17,79): use (13.16,82).

\noindent 22502: note low i; unreliable width: not used in fits.

\noindent 22535: 21 cm profile inadequate for TF use: not used in fits.

\noindent 22618: discrepancy between BMS (12.37,90) and MFB (12.16,79); use (12.26,86);
21 cm widths of BMS and MFB respectively 210 and 201: adopt 205.

\noindent 22662: note low i; also small, unreliable width: not used in fits.

\noindent 22686: width v. small: low reliability.

\noindent 22699: discrepancy between BMS (8.31,44) and MFB (8.40,47): use (8.35,46).

\noindent 22820: discrepancy between BMS (10.65,90) and MFB (10.51,77): use (10.58,86).

\noindent 22835: width v. small: low reliability.

\noindent 22863: 21cm spectrum is poor; width v. small: v. low reliability.

\noindent 22873: discrepancy between BMS (9.82,66) and MFB (9.74,63): use (9.78,65).

\noindent 22924: width v. small: low reliability; cluster membership doubtful.

\noindent 22934: width v. small, 21cm profile asymm: low reliability.

\noindent 22940: discrepancy between BMS (10.71,84) and MFB (10.64,83): use (10.67,83).

\noindent 23010: width v. small: low reliability.

\noindent 23038: 21 cm widths of BMS and MFB respectively 173 and 179: adopt 175.

\noindent 23099: width too small and poor opt. rot. curve; unreliable: not used in fits.

\noindent 23806: MFB give i=39, but their axial ratio yields i=46; we adopt i=43.
}

\vskip 0.15in\noindent {\bf Cancer:}

{\smalltype 
\noindent 4264: large galaxy, resolved by Arecibo beam; width (central beam only) uncertain; 
severe TF outlier: not used in fits.

\noindent 180141: discrepancy between HM (14.07,90) and us (13.86,86): use (13.96,86).

\noindent 180192: note low inclination: not used in fits.

\noindent 180201: Han reports type Scd, but it may as early as S0a.

\noindent 4344: note low inclination: not used in fits.

\noindent 180631: tiny disk, with nonexponential profile; photometric 
parms. highly uncertain.

\noindent 4354: discrepancy between HM (13.41,54) and us (13.65,59): use (13.53,57).

\noindent 4361: discrepancy between HM (14.00,74) and us (14.12,74): use (14.06,74).

\noindent 4375: Photometry v. uncertain: strongly affected by nearby stars in field.

\noindent 4400: discrepancy between HM (14.56,89) and us (14.91,82): use (14.73,83).

\noindent 4399: discrepancy between HM (13.64,70) and us (13.75,71): use (13.70,71).

\noindent 180270: HI spectrum very likely blend with 119-094 at 1.5'.

\noindent 4591: 21cm spectrum is blend: inaccurate extraction of width.

\noindent 4595: 21cm spectrum is poor: inaccurate extraction of width.
}

\vskip 0.15in\noindent {\bf Antlia:}

{\smalltype 
\noindent 26733: discrepancy between MFB (12.65,89) and ours (12.56,83): use (12.60,83).

\noindent 26748: discrepancy between MFB (11.85,88) and ours (12.04,81): use (11.95,83).

\noindent 26824: discrepancy between MFB (11.07,51) and ours (10.94,44): use (11.01,48).

\noindent 26943: discrepancy between MFB (12.77,56) and HM (12.57,59): use (12.66,57).

\noindent 26979: discrepancy between MFB (12.42,66) and HM (12.50,69): use (12.46,68).

\noindent 26999: discrepancy between MFB (10.79,63) and HM (10.59,69): use (10.70,66).

\noindent 27113: note v. low i; unreliable width: not used in fits.

\noindent 27146: our radial velocity (v$_{hel}$=3899) very discrepant from that of MFB 
(v$_{hel}$=2946); we adopt our velocity and width. Galaxy has LSB companion 3' NE:
possibility of blend, affecting width, cannot be excluded.

\noindent 27345: discrepancy between MFB (11.35,67) and HM (11.50,90), note inclination 
diff.; use (11.42,75), with large uncertainty.

\noindent 27358: discrepancy between MFB (10.98,59) and ours (10.88,56); use (10.94,58).

\noindent 27427: discrepancy between MFB (11.51,73) and HM (11.67,78); use (11.59,75).
Large discrepancy in widths between MFB and HM: adopt corrected width of 331$\pm$26.

\noindent 27462: discrepancy among MFB (11.89,80), HM (11.95,81) and us (11.82,78): use (11.88,78).

\noindent 27475: discrepancy between MFB (10.28,72) and ours (10.22,68): use (10.25,69).

\noindent 27495: discrepancy between MFB (11.33,81) and HM (11.49,90): use (11.41,84).

\noindent 27501: discrepancy between MFB (12.16,50) and HM (12.18,52): use (12.17,51).

\noindent 27580: discrepancy between MFB (10.69,72) and HM (10.75,71): use (10.72,71).
}

\vskip 0.15in\noindent {\bf A1060 (Hydra):}

{\smalltype  
\noindent 26740: discrepancy between MFB (12.64,83) and us (12.33,78): use (12.48,79).

\noindent 27187: discrepancy between MFB (13.18,59) and us (12.96,59): use (13.07,58).

\noindent 27374: discrepancy between MFB (13.23,70) and HM (13.31,81): use (13.29,76): 
vel discrepancy of 300 km/s!

\noindent 27404: discrepancy among MFB (12.48,75), HM (12.40,74) and us (12.14,70): 
use (12.33,72).

\noindent 27407: discrepancy among MFB (13.45,52), HM (13.17,55) and us (13.38,54): 
use (13.34,54).
                                                                                   
\noindent 27419: severe TF outlier; possible misident.: not used in fit.

\noindent 27441: discrepancy between MFB (11.77,61) and us (11.67,61): use (11.72,61). 
MFB velocity (4718) in disagreement with RC3 (5190) and ours (5243); use ours. Marginal
membership in cluster.

\noindent 27489: discrepancy among MFB (13.78,90), HM (13.79,90) and us (13.88,78): 
use (13.82,84).

\noindent 27491: discrepancy between MFB (12.19,63), HM (12.07,65): use (12.13,64).

\noindent 27529: note low i.

\noindent 27584: our opt. parms agree with those of MFB (avg parms listed), but 
MFB's rotation curve  
very poorly samples the optical disk, so width is severely underestimated. Not 
used in fits.

\noindent 27598: discrepancy between MFB (11.96,60) and HM (11.95,66): use (11.95,63).
}

\vskip 0.15in\noindent {\bf N3557 Group:}

{\smalltype  
\noindent 27705: discrepancy between MFB (11.41,80) and us (11.50,72): use (11.46,73).

\noindent 27752: discrepancy between MFB (11.24,70) and HM (11.35,78): use (11.30,74).

\noindent 27755: discrepancy between MFB (12.21,69) and HM (12.25,69): use (12.23,69).

\noindent 27756: undetected by our 140' telescope observations, which did not confirm 
Aaronson \etal (1989) Parkes detection: spurious? Not used in fits.

\noindent 27803: discrepancy between MFB (12.11,57) and HM (12.19,64): use (12.15,60).

\noindent 27829: v. large discrepancy between MFB (12.02,47) and HM (11.08,48): use (11.55,47).

\noindent 27907: discrepancy between MFB (12.07,61) and HM (12.15,62): use (12.11,61).

\noindent 27957: discrepancy between MFB (13.29,81) and HM (13.48,85): use (13.37,83).
}

\vskip 0.15in\noindent {\bf A1367:}

{\smalltype  
\noindent 6525: nearly face on; inclination unreliable: not used in fits.

\noindent 210482: uncertain morphology: bright offset nucleus, disturbed?

\noindent 6583: 21cm spectrum blend, but width recovered.

\noindent 210559: low S/N 21cm data; large uncertainty on width.

\noindent 210612: poor 21cm data; uncertainty on width large.

\noindent 6693: nearly face on; inclination uncertain: not used in fits.

\noindent 6702: discrepancy between HM (12.53,36) and us (12.69,40): use (12.61,40);
note low inclination.

\noindent 210791: discrepancy between HM (13.26,71) and us (13.12,75): use (13.18,73).

\noindent 210829: comp. within 2.2'; 21cm profile possibly confused.

\noindent 210833: 21cm profile may be blended; low quality width.

\noindent 6822: nearly face on; inclination uncertain: not used in fits.

\noindent 210859: width derived from measurement at 20\% of peak of Bothun \etal (1985).

\noindent 6876: width unreliable; possibly blend: not used in fits.

\noindent 210901: HM give mags at 23.5 mag arcsec$^{-2}$ isophote and total 
which are in great
disagreement; we assume the former is correct, apply an avg. correction to total, and 
assign large uncertainty.
}

\vskip 0.15in\noindent {\bf Ursa Major:}

{\smalltype 
\noindent 6802: uncertain mag.
}

\vskip 0.15in\noindent {\bf Cen30:}

{\smalltype 
\noindent 28431: asymm. 21cm line profile; w50 prob. underestimates width, use w20.

\noindent 28512: MFB's optical rotation curve samples small fraction of optical 
disk; width unreliable; not used in fits.

\noindent 28657: discrepancy between MFB (11.97,52) and us (11.94,51): use (11.95,51).

\noindent 28661: discrepancy between MFB (11.40,51) and us (11.35,53): use (11.37,52).

\noindent 28708: discrepancy among MFB (10.30,51), HM (10.26,52) and us (10.23,53): 
use (10.26,52).

\noindent 28756: discrepancy between MFB (13.83,83) and HM (13.73,86): use (13.78,81).

\noindent 28820: discrepancy among MFB (13.68,75), HM (13.71,77) and us (13.75,75): 
use (13.71,75).

\noindent 28832: discrepancy between MFB (12.68,60) and HM (12.68,64): use (12.68,60).

\noindent 28928: discrepancy between MFB (11.67,60) and HM (11.61,59): use (11.64,59).

\noindent 29086: discrepancy between MFB (13.87,67) and HM (13.78,74): use (13.82,70).

\noindent 29188: discrepancy between MFB (13.40,69) and HM (13.29,68): use (13.37,69).

\noindent 29277: width underest. by w50; use w=140.

\noindent 29280: discrepancy between MFB (14.68,86) and HM (14.55,90): use (14.61,85).

\noindent 29455: very low galactic latitude; galactic extinction exceeds 0.35 mag.
}

\vskip 0.15in\noindent {\bf Cen45, background and foreground of Cen30:}

{\smalltype 
\noindent 28442: discrepancy between MFB (10.34,72) and HM (10.66,75): use (10.50,73).

\noindent 28467: discrepancy between HM (11.99,68) and MFB (12.17,68): use (12.08,68).

\noindent 28696: discrepancy between MFB (13.38,79) and HM (13.59,81): use (13.48,79); 
may be Cen30 member.

\noindent 28774: discrepancy between MFB (12.58,53) and HM (12.68,60): use (12.62,57).

\noindent 28923: discrepancy between MFB (11.69,59) and us (11.74,56): use (11.72,57).

\noindent 28979: discrepancy between MFB (12.20,75) and HM (12.34,72): use (12.27,72); 
may be Cen30 member.

\noindent 29033: very low galactic latitude (13$^\circ$); large (0.51 mag), highly uncertain 
galactic extinction correction.

\noindent 29151: low galactic latitude; galactic extinction exceeds 0.4 mags.

\noindent 29153: low galactic latitude; galactic extinction exceeds 0.4 mags.

\noindent 29181: discrepancy between MFB (12.71,43) and HM (12.83,55): use (12.76,49).

\noindent 29351: discrepancy among MFB (11.80,87), HM (11.64,85) and us (11.80,80): 
use (11.75,82).

}

\vskip 0.15in\noindent {\bf A1656 (Coma):}

{\smalltype 
\noindent 7845: discrepancy between B94 (14.07,76) and us (13.91,78): use (13.99,78).

\noindent 7955: discrepancy between B94 (13.42,83) and us (13.49,83): use (13.46,83).

\noindent 221022: 21cm profile has wide pedestal; width quite uncertain: could be as wide 
as 375 or as narrow as 230. 

\noindent 7978: discrepancy between B94 (13.21,52) and us (13.15,55): use (13.18,53).

\noindent 8013: discrepancy between B94 (13.75,77) and us (13.64,79): use (13.69,78).

\noindent 8017: discrepancy between B94 (12.52,77) and us (12.62,73): use (12.57,75).

\noindent 221149: poor quality optical rotation curve.

\noindent 8082: image has very poor seeing: inclination v. uncertain.

\noindent 8096: discrepancy between HM (13.28,87) and us (13.11,90): use (13.20,88).

\noindent 8161: discrepancy among HM (13.19,64), B94 (13.14,65) and us (13.11,64): use (13.14,64).

\noindent 230051: marginal member, on caustic.

\noindent 8220: discrepancy between B94 (12.65,82) and us (12.74,85): use (12.70,83).

\noindent 8244: discrepancy between B94 (14.09,70) and us (14.16,70): use (14.13,70).
}

\vskip 0.15in\noindent {\bf ESO508:}

{\smalltype 
\noindent 29076: discrepancy between MFB (12.02,73) and HM (11.72,76): use (11.87,73).

\noindent 29082: discrepancy among MFB (12.67,89), HM (12.74,90) and us (12.71,85): use (12.69,86).

\noindent 29140: discrepancy among MFB (12.35,39), HM (12.23,48) and us (12.32,39): use (12.30,40).

\noindent 29175: discrepancy between MFB (11.75,90) and HM (11.83,90): use (11.78,88).

\noindent 29216: discrepancy between MFB (13.04,77) and HM (13.16,78): use (13.09,76).

\noindent 29264: discrepancy among MFB (12.61,90), HM (12.59,90) and us (12.52,84): use (12.56,84).

\noindent 29307: discrepancy among MFB (12.86,71), HM (12.89,80) and us (12.97,69): use (12.91,70).

\noindent 29327: discrepancy between MFB (12.71,70) and us (12.56,69): use (12.62,69).

\noindent 29332: Note low i; width unreliable. 

\noindent 530053: discrepancy between MFB (11.09,75) and us (11.13,77): use (11.11,76).

\noindent 29361: discrepancy among MFB (12.16,90), HM (12.19,90) and us (12.21,82): use 12.19,83).

\noindent 29371: discrepancy among MFB (11.62,84), HM (11.72,90) and us (11.65,81): use (11.66,82).

\noindent 29453: discrepancy between MFB (13.69,86) and us (13.73,83): use (13.71,83).

\noindent 29486: discrepancy among MFB (11.50,47), HM (11.48,42) and us (11.44,44): use (11.47,44); 
Aaronson \etal (1989) at Vhel=3066 disagree with MFB and Dressler (1991): velocity of
Aaronson \etal  probably spurious.

\noindent 29501: discrepancy between MFB (13.48,58) and HM (13.29,62): use (13.39,60).

\noindent 29511: discrepancy among MFB (12.80,88), HM (13.20,83) and us (12.72,78): use (12.90,80).

\noindent 29585: discrepancy among MFB (11.31,47), HM (11.33,47) and us (11.32,47): use (11.32,47).

\noindent 29608: feature in MFB 21cm spectrum, judged by them to be interference, may be part of signal; 
width v. uncertain and prob. underestimated. 
}

\vskip 0.15in\noindent {\bf A3574 (Klemola 27):}

{\smalltype 
\noindent 29565: discrepancy between MFB (14.14,90) and us (14.16,87): use (14.15,88).

\noindent 29708: discrepancy between MFB (10.53,74) and HM (10.36,75): use (10.45,73).

\noindent 29737: discrepancy between MFB (13.53,74) and us (13.39,72): use (13.46,72).

\noindent 29872: discrepancy between MFB (11.72,83) and us (11.77,80): use (11.75,81).

\noindent 29909: discrepancy between MFB (12.72,64) and HM (12.74,64): use (12.73,64).

\noindent 30065: discrepancy between MFB (13.17,90) and HM (13.20,90): use (13.18,84).

\noindent 30158: discrepancy among MFB (12.03,59), HM (11.93,63) and us (12.12,70): use (12.03,64).

\noindent 30307: discrepancy between MFB (11.59,51) and HM (11.34,45): use (11.47,48); note 
v. low and prob. underest. width. TF outlier by nearly 2 mags; not used in fits.

\noindent 30425: discrepancy between MFB (12.64,90) and us (12.68,87): use (12.65,87).
}

\vskip 0.15in\noindent {\bf A2197/A2199:}

{\smalltype 
\noindent 260481: marginal cluster membership.

\noindent 10389: marginal cluster membership.

\noindent 260543: very unstable isophotal ellipticity: outer isophotes have i=57, inner i=75; 
uncertainty in parms. from poorly constrained i.

\noindent 260561: note low i: not used in fits.

\noindent 260622, 10417, 10423, 260640: images taken in very poor seeing; i v. 
uncertain.
}

\vskip 0.15in\noindent {\bf S0805 (Pavo 2):}

{\smalltype 
\noindent 31805: very limited sampling of the optical disk; width v. uncertain; not used for fits.

\noindent 31871: large discrepancy between MFB optical and MFB radio widths; both spectra of 
good quality; use avg.

\noindent 31969: note low i. Rotation samples very poorly optical disk; width v. uncertain; 
not used in fits.

\noindent 32003: large discrepancy between MFB optical and MFB radio widths and velocities: 
``noise" feature in 21cm line probably part of signal; use Wopt, even if also poor.

\noindent 32005: note low i; inadequate for TF use; not used in fits.

\noindent 32068: MFB's rotation curve asymm, reaches 60" on one side, 20" on other: width and vel 
v. uncertain. 

\noindent 32175: note significant discrepancy between MFB 21cm and optical widths; both appear of
good quality: use avg.

\noindent 32709: Rotation curve samples v. poorly the optical disk; width v. uncertain; not used in fits.
}

\vskip 0.15in\noindent {\bf MDL59:}

{\smalltype 
\noindent 34434: low S/N, Gaussian--shaped 21cm spectrum; use w20=220 \kms.

\noindent 34682: discrepancy between MFB (12.37,77) and us (12.04,82): use (12.19,80).

\noindent 34685: 21cm spectrum of MFB poor; w20 appears better representation of 2$\times$Vrot 
than w50.

\noindent 34852: note low i. Inadequate for TF use; not used in fits.

\noindent 34875: note low i. Inadequate for TF use; not used in fits.

\noindent 34977: discrepancy between MFB (13.59,77) and us (13.86,78): use (13.72,78).

\noindent 34989: note width small and uncertain.

\noindent 35083: 21cm spectrum poor; after inspection, use W21=166 as better representation of 
width than 146.

\noindent 35331: discrepancy between MFB (14.47,76) and us (14.31,73): use (14.39,73).

\noindent 35781: note low inclination; not used for fits.
}

\vskip 0.15in\noindent {\bf Pegasus:}

{\smalltype 
\noindent 12382: discrepancy among HM (14.43,90), MFB (14.37,81) and us (14.40,80): 
use (14.40,80); TF outlier by 1.5 mag: not used in fits.

\noindent 12417: discrepancy among HM (14.03,67), MFB (12.76,61) and us (12.72,62); 
prob. misidentification by HM: use (12.72,62).

\noindent 12423: discrepancy among HM (12.39,90), MFB (12.57,90) and us (12.30,81): 
use (12.40,83).

\noindent 12437: discrepancy between HM (12.24,66) and MFB (12.37,65): use (12.30,65).

\noindent 12467: discrepancy between HM (13.30,79) and MFB (13.75,81): use (13.52,79).

\noindent 12483: discrepancy between HM (12.66,50) and MFB (12.76,57): use (12.71,54).

\noindent 12494: discrepancy between HM (13.68,74) and us (13.93,77): use (13.80,76); 
also note that 12494 is in double system.

\noindent 12498: discrepancy between HM (12.37,62) and us (12.63,62): use (12.50,62).

\noindent 330261: i highly uncertain: not used in fits.

\noindent 12555: discrepancy among HM (14.46,79), MFB (14.75,78) and us (14.75,77): 
use (14.65,77).

\noindent 12561: discrepancy between HM (13.92,77) and MFB (14.41,75): use (14.16,75).
}

\vskip 0.15in\noindent {\bf A2634/A2666:}

{\smalltype 
\noindent 330564: marginal cluster  membership assignment.

\noindent 331234: in probable interacting system: cautionary TF use.

\noindent 330663: poor (S/N) on spectrum; v. uncertain width.

\noindent 12721: discrepancy between HM (13.01,63) and us (12.76,65): use (12.89,64).
}

\section {Inclinations}

For galaxies for which we had available access to the I band images, 
inclinations were obtained
after fitting ellipses to isophotal contours using the ISOPHOTE routine in
STSDAS, as described in paper I. From these isophotal fits, one can obtain
an ellipticity  --- defined as $e = 1- b/a$, where $a$ and $b$ are 
the major and minor axes of the ellipse --- as a function of the distance 
$r=a/2$ from the center of the galaxy. A range of radial distances is then
chosen over which the disk appears to be exponential, and a mean value of
the ellipticity $\bar e$ is obtained. This parameter yields a first
order approximation to the true ellipticity of the image, which needs to
be corrected for the effects of resolution, mainly the smearing effects of
seeing. We assume that the disk light drops to half power near 
$a' = 0.2 a_{23.5}$ and $b' = 0.2 b_{23.5} = 0.2a_{23.5}(1-\bar e)$ 
respectively along the major
and minor axes, where $a_{23.5}$ and $b_{23.5}$ are the major and minor
isophotal sizes at the 23.5 I mag sec$^{-2}$ isophote; this results from
the fact that, statistically, $a_{23.5}$ is about equal to 3.5 scale lengths.
Then if $\psi = 0.5 {\rm HPFW}_{seeing}/a'$, where HPFW$_{seeing}$ is the size 
of the seeing disk for the given observation, we adopt as the corrected 
ellipticity

$$e_{corr} = 1 - \sqrt{{(1-\bar e)^2 - \psi^2}\over{1 - \psi^2}} \eqno (2)$$

\noindent
In order to avoid overcorrection, we limit 
$e_{corr}$ to not exceed a maximum value of $(0.12)^{-1}$, and 
assume $e_{corr}$ is a good approximation of the true value of $e$.
In the following, when we refer to $e$, we mean the corrected value
given by eqn. (2), unless otherwise specified.

The inclination angle is then derived as

$$\cos^2 i = {(1-e_{corr})^2 - q_\circ^2 \over {1 - q_\circ^2}} \eqno (3)$$

\noindent
where $q_\circ$ is the intrinsic axial ratio of spiral disks, which are treated
as highly oblate spheroids. For Sbc and Sc galaxies, we adopt $q_\circ=0.13$, as
discussed in paper I; for other types we adopt a more conservative $q_\circ=0.2$.

In the published literature, inclinations have occasionally been obtained
using criteria different from our own. Both MFB and HM, for example, adopt
an intrinsic disk axial ratio of 0.2 for all types. In those cases, we
rederive an axial ratio from the published inclination, and use it to
estimate the inclination according to our choice of intrinsic axial ratio $q_\circ$.
A problem with this approach is that published inclinations obtained
using $q_\circ=0.2$ yield values of $e$ that saturate at 0.80; for a
$q_\circ=0.13$, that ellipticity yields an inclination of $81^\circ$. In the
case of MFB, in addition to inclinations also axial ratios are given, so
that the ambiguity can be partially solved: whenever their published inclination
is $90^\circ$ (which would yield $81^\circ$ for a $q_\circ=0.13$), we recompute
the ellipticity from their reported axial ratio, and use it to estimate
the inclination, with the value of $q_\circ$ appropriate to our convention.
In the case of HM's data, axial ratios are not available, and inclinations
reported at $90^\circ$ are kept at that value, even if for a small fraction
of objects such inclination may be a slight overestimate. No seeing corrections 
are attempted for data other than our own.

Uncertainty in the determination of the disk inclination propagates
to variables on both axes of the TF diagram. The correction to magnitude
for internal extinction increases with increasing inclination angle, while
the correction to the observed velocity widths is largest for systems
closer to face--on. As ellipses were fitted to each galaxy image, we 
estimated errors in the determination
of $e$ for our data, which include not only the cluster galaxies
listed in Table 2 but also a larger field sample of late--type spirals. 
In figure 13, we display the errors $\epsilon_e$, as derived from our
data, plotted versus $e$. A trend is discernible in the sense that
$\epsilon_e$ increases with decreasing $e$. 
The solid line superimposed on the data 

$${\rm med}~\epsilon_e = 0.090 - 0.12e + 0.037 e^2 \eqno (4)$$

\noindent
yields an estimate of the expected (median) error associated with the 
ellipticity. We can use eqn. (4) in the computation of the
error $\epsilon_e$ for sources of data other than ours, for which 
values of the ellipticity or axial ratio are given without an indication 
of the associated error.

The distribution of inclination errors shows considerable skewness and kurtosis,
especially for low values of the inclination. This combined effect can
arise from isophotal maps strongly affected by the light of a bar, the 
presence of an undetected warp, or other disk distortions.

\section {Velocity Widths}

The velocity widths have been obtained at several different telescopes, 
using both radio and optical long--slit techniques. All the radio spectra
have been reduced within the same system. Techniques and algorithms
applied in the reduction of our data, as well as comparisons with other 
data sets are described elsewhere (Haynes \etal 1997). Here we are
mainly concerned with the details related with the characterization of
the velocity--width error budget.

\subsection {Corrections Applied to 21 cm Velocity Widths}

Our 21 cm velocity widths were measured using an algorithm that fits the 
rising sides of the galaxian profile with low order polynomials
and obtains the width at a 50\% level of each of the profile horn peaks 
or single peak, depending on the line shape. Similar techniques have been 
applied in data drawn from other sources, and we have applied each of
the standard techniques to our data as well, and derived recursion
relations to convert widths from other sources to our standard. In order 
to derive a measure of the disk maximum rotational velocity
from the observed 21 cm profile width, a number of corrections need
to be applied. 

First, the broadening effects produced by the instrumental characteristics,
the signal--to--noise ratio (S/N) of the spectrum and any smoothing that may 
have been applied in order to improve the (S/N) need to be accounted for. 
These broadening processes depend on the shape of the spectral line, i.e.
the steepness of the ``horns''. 

Second, both turbulent and ordered motions contribute to the profile
width. We assume that ordered motions are fully produced by rotation,
and attempt no corrections related to large scale phenomena such
as warps of the disk, noncircular orbits and motions perpendicular to
the disk plane. We statistically correct only for the contributions
of turbulent motions.

Third, the observed profile is broadened in the measure $(1+z)$, where 
$z$ is the redshift of the source, by relativistic effects.

Fourth, the Doppler observations refer to the line of sight projections
of the gas motions. The amplitude of ordered motions occurring in the 
plane of the disk can be obtained from the observed values when the
inclination angle $i$, between the normal to the plane of the disk 
and the line of sight, is known.

The {\it corrected} velocity width will then be

$$W_c = \Bigl[\Bigl({W_{obs,21} - \Delta_s \over {1+z}}\Bigr)^2 - \Delta^2_t \Bigr]^{1/2}
{1\over{\sin i}} \eqno (5)$$

\noindent
where $W_{obs,21}$ is the observed 21 cm width and $\Delta_s$ is the 
correction for the effects of instrumental broadening and of any smoothing 
applied in the reduction stage. The latter term also takes into account 
the distortions that result with varying (S/N) and a dependence on the 
observed shape of the line profile. The term $\Delta_t$ accounts for the 
broadening produced by turbulent motions, and the factor $1/\sin i$ 
deprojects the disk.

\subsection {The Velocity Width Error Budget}

Each one of the correction steps adds a measure of uncertainty to the 
determination of the velocity width, which needs to be carefully tracked 
since, as we shall see, they constitute one of the principal contributions 
to scatter in the TF diagram.

Let the measurement of the spectrum yield a width $W_{obs,21}$, with an
associated error on the measurement $\epsilon_{w,obs}$.

Based on our own numerical simulations, we assume that the uncertainty 
associated with the quantities $\Delta_s$ and $\Delta_t$ is approximately
25\% of the value of each, i.e. $\epsilon_{s} = 0.25 \Delta_s$, and
$\epsilon_{t} = 0.25 \Delta_t$.

The uncertainty on $1/\sin i$, the inclination correction, depends
principally on the accuracy with which the disk ellipticity has been 
measured. The dependence of $\epsilon_e$ on $e$ is shown
in Figure 13 and parametrized by eqn. (4). The error on the measurement of 
$e$ propagates to that in $1/\sin i$ via

$$\epsilon_i^2 = \Bigl[{d(1/\sin i)\over {d(e)}}\Bigr]^2 \epsilon_e^2 
\eqno (6)$$

\noindent
where we have neglected the effect of the uncertainty on the value of $q_\circ$.

The uncertainty $\epsilon_w$ associated with $W_c$ is then

$$\epsilon^2_w = {\epsilon^2_{w,obs} + \epsilon^2_s\over {(1+z)^2\sin^2 i}}
+ {\epsilon^2_t\over {\sin^2 i}}
+ \epsilon^2_i W_c^2\sin^2 i      \eqno (7)$$

\noindent
The relative error on the width, $\epsilon_w/W_c$, increases 
with decreasing inclination, because the first two terms in eqn. (7)
are proportional to $1/\sin^2 i$ and because $\epsilon_i\sin i$ increases 
as the disk gets closer to face--on, as shown in Figure 13. Similarly,
$\epsilon_w/W_c$ increases with decreasing $W_c$,  because $\epsilon_{w,obs}$,
$\epsilon_{s}$ and $\epsilon_{t}$ are to the first order independent of width, 
and therefore constitute an increasingly large fraction of the width as
$W_c$ decreases. This is shown in Figure 14.

The relative importance of the various terms contributing to $\epsilon_w$
does of course depend on the inclination of the system. For highly 
inclined systems, the uncertainty on $i$ does not play a very important
role, while for systems with $i<65^\circ$, the term in $\epsilon_i$ can be 
the most important. The terms in $\epsilon_s$ and $\epsilon_t$ seldom account
for more than a quarter of the total error budget, and more often they
account for less than 10\% ; in addition, in most cases in which these 
terms account for more than 10\% of the width error, the latter 
is usually quite small in absolute terms. Thus, the simplicity of our 
approximation of $\epsilon_s$ and $\epsilon_t$ as constant fractions of the 
corrections $\Delta_s$ and $\Delta_t$, respectively, is well justified; 
a more elaborate parametrization of $\epsilon_s$ and $\epsilon_t$ is unnecessary.

\subsection {Corrections applied to Optical Velocity Widths}

For a sizable fraction of objects in our sample, we use velocity widths 
obtained from optical, single slit spectroscopy. For these objects, adequate 
21cm line profiles are generally unavailable. A comparison of optical and radio widths,
and a detailed discussion of the derivation and treatment of both quantities
is given elsewhere (Haynes \etal 1997).

A number of corrections need to be applied to optical velocity widths;
some are similar to those applied to radio widths and described above,
others are not, such as those responding to the {\it ad hoc} requisite that 
the corrected widths
be compatible with those obtained via radio means.

First we apply an additive offset $\Delta_{sh}$ to the observed or
reported optical width, 
$W_{obs,opt}$, such that this correction will make it equivalent to the radio 
width, corrected for instrumental, smoothing and signal--to--noise broadening 
effects. Such offset depends principally on the adopted measuring technique 
and on the shape of the rotation curve, as discussed in Haynes \etal (1997). 
Because the HI 
widths tend to sample regions of disks that generally extend beyond the edge 
of the regions sampled by optical rotation curves, the offset between the two 
widths depends on the extent over which the spectrum spatially maps the
disk, and on whether the rotation curve is seen to be still rising, flat, 
or falling at the outer parts of the disk sampled by the spectrum. 
In a reanalysis of the optical rotation curve data of MFB, Persic and 
Salucci (1996) have adopted a velocity width measurement that corresponds
to the value of the rotational speed at a radius $R_{opt}$ within which 
80\% of the
disk's I band light is enclosed. We also adopt here that criterion.
For objects for which both optical rotation curve and 21 cm line data
are available, the corrected 21 cm width and the optical width measured
at $R_{opt}$ provide a good match. For galaxies
where the sampling of the optical spectrum does not reach as far out
as $R_{opt}$, we extrapolate the rotation curve, using the parametrized
``universal rotation curve'' form of Persic and Salucci (1991), as 
described in detail in Haynes \etal (1997). A second offset correction
would be necessary, in order to account for poor estimates of the position
angle (PA) of disks' major axes. This should always be an additive term,
because any misestimate of the PA will generally reduce the measured width.
An estimate of the error incurred in estimating the PA is difficult. We
have simulated the effects of errors on the observed width by misalignment
of the spectrograph slit, both in PA and in centering, and have found that
the errors remain negligible for $PA<15^\circ$ or so. Given the uncertainty
in estimating the PA error, we chose not to apply a correction for this
effect. Any net mean offset would be small, and would be incorporated in
the $\Delta_{sh}$ term.

Next, a $(1+z)^{-1}$ relativistic correction and  an inclination correction 
are applied, as described for the radio widths in the preceding section.
The corrected width derived from optical rotation curves is then

$$W_c = {{W_{obs,opt} - \Delta_{sh}}\over {(1+z)\sin i}}, \eqno (8)$$

\noindent
and the resulting uncertainty associated with $W_c$ is 

$$\epsilon^2_w = {\epsilon^2_{w,obs} + \epsilon^2_{sh}\over {(1+z)^2\sin^2 i}}
+ \epsilon^2_i W_c^2\sin^2 i  \eqno (9)$$

\noindent
where $\epsilon_{w,obs}$ is the measurement uncertainty on $W_{obs,opt}$,
$\epsilon_{sh}$ is the uncertainty on the shape dependent offset correction
$\Delta_{sh}$, the amplitude of which is very sensitive on whether $W_{obs,opt}$ 
is obtained from
the rotation curve by interpolating between observed values or extrapolating
beyond the last measured radius;  
$\epsilon_i$ is the uncertainty on $(1/\sin i)$, as shown in the preceding
section. For the widths derived from the data of MFB, we have inferred
uncertainties $\epsilon_{w,obs}$ from the tabulation of folded and 
kinematically recentered
rotation curves by Persic and Salucci (1996). When we use data from sources 
other than ours, for which
$\epsilon_{w,obs}$ is not reported and for which a more informed estimate of the
width error is not possible, we assume $\epsilon_{w,obs}=20$ \kms, an error
typical for galaxies with measured errors.

\section {I band Fluxes and Related Errors}

For all the objects used in this study, magnitudes derived from
I band CCD images are available. In some cases, other photometric parameters, 
such as scale lengths and isophotal sizes, are also available. When scale 
lengths are measured, it is possible to infer an asymptotic magnitude, which 
results from extrapolating the ``curve of growth'' of isophotal
magnitudes to infinity, assuming that the surface brightness falls 
exponentially with the measured scale length. Such extrapolation is
applied to the magnitude measured at an isophotal level between 23.5 and 
24.0 mag arcsec$^{-1}$ (which is found at a distance of $~4$ scale lengths 
from the center of the galaxy), and it amounts typically to a few hundreds 
of a magnitude. We adopt such an asymptotal measure as our estimate of the 
I band flux.

There is a significant offset in the mean values of the total uncorrected 
magnitude, between our own observations and those from the other main
source utilized in this paper, i.e. those by MFB. We compare
217 galaxies for which both MFB and our magnitudes are available. Of those, 
196 have magnitude differences smaller than 0.3 magnitudes; galaxies with
magnitude discrepancies larger than 0.3 mag are not used further in comparing
the two scales. The 196 galaxies yield a mean difference in the measured 
total magnitude of $<m_{our} - m_{MFB}> = -0.053$. This difference does however
exhibit a systematic trend, in the sense that the difference becomes larger
(more negative) for objects of brighter apparent magnitude. We tentatively
attribute part of the effect to the fact that some of the observations of
MFB of objects of relatively large angular size were carried out with a
relatively small chip on the 1m telescope at Siding Spring; sky subtraction
may have been less effective than for the larger objects observed with that
system. We apply a correction to the MFB magnitude scale as follows:

$$m_{MFB}^{cor} = m_{MFB} - 0.08 - 0.016\delta + 0.011\delta^2 \eqno (10)$$

\noindent
for objects of $m_{MFB}<13.6$, where $\delta = m_{MFB}-10$, and 
$m_{MFB}^{cor} = m_{MFB}$ for $m_{MFB}\ge 13.6$.

The overlap of objects between our and the HM samples (51) and that between the 
HM and MFB samples (86) are too small to yield a statistically significant
offset between scales. We thus use the raw HM magnitudes unchanged. 
The scatter in the comparisons gives an idea of the reliability of I band
magnitudes. While comparisons of repeat observations within a given data set 
report consistency to 
within 0.04 mag, comparison between different data sets suggests that
the quality of the data may be slightly worse, on the order of 0.06--0.07 mag.
Large discrepancies, in excess of 0.15 mag, occur more frequently than 
would be expected if errors were random; these cases probably arise from
the occasional misjudgement of photometric conditions, or from 
misidentification of the source. In the tabulation of data utilized for 
this study, we footnote large discrepancies between different data sources. 
When several sources are available, we generally adopt average values, 
else we justify our choice in the notes to Table 2.

Adopted raw magnitudes are corrected first for extinction within our Galaxy,
by adopting the Burstein and Heiles (1978) tabulation of B--band extinctions,
converted to I--band by means of $A_I = 0.45 A_B$, and by a cosmological k--term, 
adopted from Han (1992): $ k_I = 0.16 z$.
The internal extinction in each galaxy's disk is treated in the manner described
in paper II. All magnitudes are converted to a face--on perspective, by applying
an additive term which depends on both the inclination of the system to the line
of sight and on its intrinsic luminosity.
The corrected apparent magnitude is then

$$m_c = m_{obs} - A_I + k_I - \Delta m_i  \eqno (11)$$

\noindent
where $m_{obs}$ is the raw asymptotic magnitude and $\Delta m_i$ the internal 
extinction correction. $m_{obs}, ~A_I, ~k_I, ~\Delta m_i$ are positive quantities. 
The internal extinction correction was parametrized in paper II as

$$\Delta m_i = - \gamma \log (1 - e) = \gamma \log (a/b) \eqno (12)$$

\noindent
where $e$ is the corrected ellipticity and the parameter $\gamma$ 
depends on absolute luminosity of the galaxy as shown in Figure 7c of paper II
($\gamma = 0.5$ for $M_I > -19$, $\gamma = 1.0 $ for $-20.3 > M_I > -22$,  
and $\gamma$ changes linearly by -0.4 mag$^{-1}$ elsewhere).

For application in a TF program, however, it is convenient to have the value 
of $\gamma$ parametrized as a function of velocity width. In Figure 15,
we plot the magnitude residuals with respect to a direct fit to the 
template TF relation of the cluster sample, which we will obtain in 
Paper VII. The data plotted pertain to the all sky spiral sample
described in Papers I and II  (filled circles) and the cluster sample 
(unfilled squares). The magnitude residuals are computed after correcting 
the magnitudes by imposing $\Delta m_i = 0$ in eqn. (11), 
and separately for five different ranges of velocity widths;
since residuals are plotted versus $\log (a/b)$, their trends mimic the 
internal extinction correction that should be applicable. 

The solid lines inset on panels (c) through (e) are not fits 
to the data points but rather the mean internal extinction laws that would apply
to the subset of galaxies, according to the model of eqn. (12) derived
for late--type spirals (the slope
of the line is the value of $\gamma$ which would be used, according to
the precepts of paper II, averaged over all the galaxies in the subset).
We adopt those lines as our internal extinction laws. For the fastest rotators 
the internal extinction appears to  
increase slowly for values of $\log (a/b)$ smaller than about 0.5, then 
more steeply at higher inclinations. The apparent bilinearity
of the internal extinction law may be related to a decrease of scale height
to scale length ratio with overall galaxy size, or to a decreasing degree of
clumpiness of opaque regions in the disk with increasing disk size: both
effects would shift the inclinations at which internal extinction increases
more steeply to values closer to edge--on. 
The departure of the data from the simple model of eqn. (12) is significant
enough, especially for objects of intermediate inclination in panel (a) of 
Figure 15, that the adoption of an alternative model is warranted. In order 
to keep it simple, for the fastest rotators we adopt a bilinear model, which
involves two different values of $\gamma$, respectively for high and low
values of $\log (a/b)$. The adopted values of 
$\gamma$, tabulated for different values of velocity width, are given in 
Table 3; for the two bins of higher width, two values of $\gamma$ are given,
meant to apply respectively below and above the value $\log (a/b) =0.52$.
Table 3 is used in the estimate of the internal extinction correction that 
leads to the values listed in col (14) of Table 2. Finally, for objects with 
$i=90^\circ$, an additional 0.05 mag internal extinction term is adopted.
Higher extinction occurring in edge--on systems is to be expected from
extinction models, and visually dramatized by the effect of dust lanes
bisecting disks; it is also statistically suggestive in our data, which 
includes 31 objects listed at $i=90^\circ$. A correction
of only 0.05 mag over the inclination dependent law may be conservative,
but it is likely that a significant fraction of the systems deemed to
be at $i=90^\circ$ are not exactly edge--on (the determined value of $i$
being highly sensitivity on the assumed value of $q_\circ$, see eqn 2), and 
additional 0.05 mag of extinction is a fair statistical correction which 
averages over such systems.

The cluster sample contains a sizable fraction of galaxies of type earlier
than Sbc, for which all our extinction relations had been obtained. On the
grounds of suggestive statistical evidence, based on the comparison of 
about 30 galaxies earlier than Sab and about 160 galaxies of type Sb with
the galaxies of later types, we adopt for all galaxies of type 
earlier than Sbc an extinction relation which, at any given inclination,
is 85\% of that estimated for the Sbc and Sc galaxies as shown in Figure 15 
and Table 3. This ``differential treatment'', which is typically smaller 
than the total error on the magnitude, as discussed below, has a negligible
impact on the final template obtained in paper VII. While it needs to be 
corroborated by the study of larger samples, the sense of the correction
appears clear, and it points in the 
direction of an increasing ratio between scale heights of the dust and of 
the stellar population sampled by the I band, towards later types. 

Estimates of the accuracy of our tabulated magnitudes derive from the
addition in quadrature of two main quantities. One is the measurement
error, resulting from estimates of the photometric conditions at the
time of data taking, the quality of the sky subtraction, the effects
of neighboring interfering objects in the frame, the quality of the
asymptotic extrapolation of the light profile, etc. In general, these
sources of errors are not itemized in the literature. Whenever an
estimate of this error is not given in the sources used other than
our own, we assume a value of 0.04 mag. If more than one independent
measurement of the magnitude is available for a given object, we adopt
as a measurement error either 0.04 mag or one half of the magnitude
difference between extreme values, whichever is largest. The second source
of error derives from the corrections expressed by eqn. (11), of
which $\Delta m_i$ is usually the most important. We adopt an
uncertainty of 15\% on the value of $\gamma$ and propagate the error,
together with that on the uncertainty on the axial ratio, to that on
the estimate of the extinction correction. This component of the 
uncertainty typically exceeds the measurement error, for inclinations 
larger than about 60$^\circ$.  In paper VII, we shall refer to the
combination in quadrature of measurement and processing errors described
above as the ``total measurement error'' on the magnitude.

\section{Global Characteristics of the Cluster Sample}

Here we investigate the global properties of the 
cluster galaxies listed in Table 2. As discussed in section 3, for each 
cluster, two principal subsamples have been defined: one of objects believed
to be cluster members ({\bf in}) and a second ({\bf in+}) that, in addition  
to cluster members, also includes objects in the periphery and near the 
systemic velocity of the cluster, but sufficiently removed (extending to 
about 3 or 4 Abell radii) that a membership assignment cannot be confidently 
made.

The distribution of apparent I band magnitudes of all galaxies in the
{\bf in+} samples is shown in Figure 16. It illustrates that for this 
sample no assumption of completeness to any useful flux limit 
can be made. Similarly, Figure 17 displays the distribution of isophotal 
radii of the same galaxies, as measured at $\mu_I = 23.5$ mag arcsec$^{-2}$,  
demonstrating a similar unsuitability of the data to be usefully 
analyzed with criteria applicable to angular size limited samples.

Many haphazard circumstances affect the construction of a cluster TF sample, 
which eventually lead to distributions such as those in Figures 16 and 17: 
inaccuracy of parameters listed in catalogs from which samples are extracted, 
a priori availability of redshifts, telescope scheduling, weather conditions, 
a posteriori sifting by final data quality, availability of data from 
external sources based on different selection criteria, etc. The analysis
of the data requires that corrections be applied, that take into consideration
the peculiarities of the sample selection. Unfortunately, 
individual cluster samples are still small enough that sacrifice of a fraction 
of available objects made to obtain a clean subset of simple completeness 
characteristics, and to which straightforward corrections can be applied, is 
painfully impractical. In order to maintain sample integrity and overcome the 
effects of selection biases, we believe the best current approach is that of 
estimating corrections via computer simulations with artificial samples that 
reproduce the completeness characteristics of the data.

An analysis of the completeness characteristics of a larger sample of spirals, 
which includes the current cluster sample as a subset, is presented in Papers
I and II. In Paper II, we used visibility functions of luminosity and 
central disk surface brightness to illustrate built--in biases in the optical
catalogs from which the observed objects were extracted. 

In Figure 18, we show the distribution of absolute magnitudes of all the
galaxies included in the {\bf in+} cluster samples. The line superimposed
onto the histogram represents our best estimate of the spiral galaxy 
luminosity function (LF) from which the observed sample may be assumed to have
been extracted. The latter is parametrized by a Schechter formula with
$M^* + 5 \log h = -21.6$ and $\alpha = -0.50$, obtained by a match between
the observed distribution in each cluster with the luminosity functions 
of spiral galaxies derived by Sandage, Binggeli and Tammann (1984) in their 
survey of the Virgo cluster. The skewness of the adopted 
LF departs from the nearly gaussian luminosity function of 
normal spirals found in the Las Campanas survey of Sandage \etal (1984). 
This is due to a small
contribution of Sdm systems which contaminates the ``normal'' spiral sample. 
Objects fainter than $M_I\simeq -19$ are only observed in the nearby
clusters/groups, such as Fornax, Eridanus, Ursa Major and MDL59.
For comparison, the Milky Way galaxy has an approximate I band absolute
magnitude of -22, M31 of -22.5 and M33 of -19.7 (Pierce and Tully 1992).
The average color (B-I) of spirals such as those in our sample is about
1.5 mag. The shape of the LF from which cluster samples
are extracted is important in the computation of incompleteness bias
corrections. In Paper VII we carry out those computations for two values
of $\alpha$, including the one quoted above, in order to ascertain the
sensitivity of the correction on the characteristics of the adopted LF.

\section {Membership Assignment}

Cluster membership was assigned according to criteria that varied somewhat from
cluster to cluster, depending on how well the cluster parameters are known.
In clusters for which a large body of redshift information is available, and
for which reliable estimates of the velocity dispersion and caustic curves
are known, membership assignment has been extended with some confidence to
larger distances from the cluster center. On the other hand, for less well
defined clusters, a more conservative approach has been adopted:
membership was restricted to objects projected more closely to the cluster 
center and/or with velocities closer to its systemic velocity.

In some cases, clusters are located at small angular and velocity separations
from nearby aggregates, such as in the case of the NGC 383 and NGC 507 groups,
Fornax and Eridanus, A2197 and A2199, A2634 and A2666. In each case, 
assignment to the {\bf in} and {\bf in+} samples was made according
to criteria of angular and velocity separation, similar to those extensively
described in the study of the A2634/A2666 system by Scodeggio \etal (1995).
In some cases, galaxies in cluster peripheries may with nearly equal likelihood
be considered peripheral members of either of two nearby clusters; we have
then avoided entry duplication, assigning each galaxy to only one cluster
listing. A discussion of choices for individual cases follow.

{\it NGC 383 and NGC 507 Groups. } The centers of the two groups are separated
by $3.4^\circ$ and are virtually at the same redshift (Sakai \etal 1994). The
two clusters have often been lumped into a single unit referred to as the 
``Pisces cluster'', although Sakai \etal are able to clearly separate their 
binary nature. As a visual separation criterion one can adopt the line of 
RA = 01$^h$ 10$^m$. There is no ambiguity in the membership assignment to any 
objects projected inside 
the circle of 1 $R_A$ radius. Three galaxies projected between the circles of
1 $R_A$ and 2 $R_A$, UGC 764, UGC 810 and UGC 820, are assigned to NGC 507,
although they are nearly equidistant to the center of the NGC 383 group. A
percolation algorithm assigns them to NGC 507, which appears to be the more
conspicuous of the two groups. An effect of the proximity of the two groups
is that the peripheral members listed in the NGC 383 table are preferentially
to the West of the cluster center. A262 is located approximately $7^\circ$
to the NE of the center of NGC 507, bounding the group to the east.

{\it Fornax and Eridanus. } Eridanus is a group about 14$^\circ$ to the North
of the well developed Fornax cluster. It is listed as group G31 by de
Vaucouleurs (1975), and its center is often associated with the location of 
the S0 galaxy NGC 1332. An appraisal of the distribution of galaxies in the 
region does however suggest that this loose aggregate is organized around 
several high ranked but well separated galaxies. We have adopted as the center 
a position intermediate between the locations of the bright galaxies NGC 1325, 
NGC 1332, NGC 1353, NGC 1359, NGC 1395 and NGC 1407, i.e. RA = 03$^h$ 30$^m$, 
Dec = $-21^\circ$ 30'. The distribution of redshifts of 148 galaxies with 
$cz < 3000$ 
\kms, within 13$^\circ$ from that center and North of Dec = $-28^\circ$ (a
boundary chosen in order to avoid confusion with Fornax), looks fairly regular 
and centered near $V_{cmb} = 1530$ \kms, which is very close to the central 
velocity adopted by MFB of 1534 \kms . There is no overlap of the areas 
enclosed within the respective 1 $R_A$ circles of each cluster. Galaxies
416-G6, 416-G20 and 486-G5, which are roughly equidistant from the centers 
of both Fornax and Eridanus and at about $2R_A$ from each, are assigned as 
peripheral members ({\bf in+} sample) to Fornax; this decision does not affect 
their location in the TF diagram, as distances are assigned to them on the 
basis of their individual redshifts, rather than that of either cluster.

{\it Cancer. } The structure of this cluster has been studied in detail by 
Bothun \etal (1983), who identified several groups, named A through E, spread in
velocity over 2800 \kms. In Table 2 we have assigned membership to each group
according to the criteria of Bothun \etal, and label as cluster members those
objects pertaining to the largest of the groups, group A. In Figure 5, members
of the foreground group D are identified by unfilled triangles, while members
of background groups B and C are identified by unfilled squares. For all 
galaxies, angular distances from the cluster center, $\theta$, are referred 
to the center of group A.

{\it A1367. } Membership assignment to A1367 has been extended to about 
4$^\circ$ from the center, well in excess of 2 $R_A$, because the 
cluster structure is well defined by a rich redshift data base, there are no 
confusing nearby clusters or groups, and caustic curves have been reliably 
determined by Reg\" os and Geller (1989). We interpolate their set of caustic 
curves to obtain those corresponding to a value of the density 
parameter $\Omega_\circ = 0.3$, and use the latter to define cluster membership.

{\it Centaurus. } Lucey \etal (1986) give a detailed analysis of the structure
of the Centaurus cluster and are able to separate two main components, both
centered at the same sky location but with redshifts differing by about 1500 
\kms. The two clumps are known as Cen30 and Cen45. In Figure 8, galaxies assigned 
to Cen45 are identified by unfilled squares, while those assigned 
to the more conspicuous Cen30 cluster are identified by filled circles. The 
separation between the two groups, on a galaxy per galaxy basis, is tenuous. 
Some of the objects assigned to Cen45 on the basis of their radial velocity 
may be ``redshift outliers'' of Cen 30. This case is more
likely for 322-G48 and 323-G42; in trying to keep the Cen30 sample as
free of contamination as possible we have conservatively 
included them in Cen45. Since only 6 galaxies are assigned membership to 
Cen45, and because of the uncertainties in the assignment, we deem incautious 
an estimate of the mean distance of Cen45 using such a sparse and uncertain 
sample. We notice that in the periphery of the cluster, at $\theta > 3^\circ$,
the mean redshift of galaxies with known redshift is perhaps 160 \kms larger 
than the systemic velocity adopted for Cen30 (as derived by Lucey \etal). 
Given the difficulty in separating Cen30 and Cen45, the possibility can 
be suggested that the true velocity of Cen30 may be slightly higher than 
estimated by Lucey \etal and adopted here. This unusually large uncertainty 
on the systemic velocity of the cluster should be folded in any estimate of 
the uncertainty of its peculiar velocity.

It should be noted that the Centaurus region lies at very low galactic
latitude, and the corrections for extinction arising in the Milky Way are
important, occasionally exceeding 0.3 mags. Because galactic extinction
is quite patchy, those corrections can be rather uncertain, especially
when their amplitude is large. This accounts for the significantly
larger magnitude corrections in this region,
and may be an important source of the large scatter in the TF diagram
for this cluster (see Paper VII).

{\it Coma (A1656). } A situation similar to that discussed for A1367 is 
encountered here. The cluster has been very well studied and reliable caustic 
curves have been published by Reg\" os and Geller (1989). We extend membership 
$2 R_A$, to galaxies beyond included within the $\Omega_\circ = 0.3$ caustic 
curves, out to a maximum radial distance of $\theta = 4^\circ$. This is 
possible because the cluster structure is well sampled and no confusing clumps 
appear projected on it. The substructure reported in West \etal (1995, and 
refs. therein) in Coma is all enclosed within the inner 1 $R_A$ of the cluster,
and does not raise concerns regarding membership assignment for the spiral
population.

{\it A2197/A2199. } The two clusters are separated by less than $1.5^\circ$ and 
200 \kms, respectively on the sky and in systemic velocity . Separation of the 
two clusters on a galaxy per galaxy basis, outside the central cores, is 
uncertain. Nonetheless, the mistaken assignment of a galaxy to one cluster
rather than the other in the binary system would produce an error in the 
estimated magnitude of less than 0.05 mag, comparable with the 
uncertainty that derives from our measurement and inclination corrections. The 
two galaxies for which membership assignment is most doubtful are 260930 and 
260561, which have been tentatively asigned to A2197; 250561 is not used 
in TF fits, however, because of its low disk inclination of $37^\circ$.
The impact of cluster membership concerns is thus relatively minor
in this region.

{\it Pavo I and Pavo II.} These are two rather poorly defined, sparse groups 
with a comparatively small number of available redshifts. We have obtained 
approximate center coordinates and systemic velocities using the available
redshift data base. In doing so, we have set an arbitrary boundary between
the domains of the two groups as follows: Pavo I is at $Dec < -68^\circ$ for
$19^h < RA < 19.8^h$, and $Dec < -65^\circ$ for $RA > 19.8^h$. Within that
rough division, we identify Pavo I based on 32 redshifts at $cz < 6000$ \kms
and within $5^\circ$ of RA = 20$^h$ 13$^m$, Dec = $-71^\circ$ 00', to which
we assign a systemic velocity of $V_{cmb} = 4104$ \kms . The uncertainty of 
this determination is no better than 100 \kms . For comparison, note that
MFB assume a systemic velocity of $V_{cmb} = 3518$ \kms. For Pavo II the 
data base is better; using 90 redshifts with $cz < 6000$ \kms and
within $5^\circ$ of RA = 18$^h$ 42$^m$, Dec = $-63^\circ$ 20', we assign to
Pavo II a systemic velocity  $V_{cmb} = 4470$ \kms, with an uncertainty on 
the order of 70 \kms and a velocity dispersion of about 450 \kms. For 
comparison, Bell and Whitmore (1989) obtain a systemic velocity of 4437 \kms 
and a velocity dispersion of 507 \kms, based on 42 redshifts. MFB distinguish
between Pavo II and DC1842--63, assigning $V_{cmb} = 4712$ to DC1842--63
and $V_{cmb} = 4357$ to Pavo II, which they identify with a sparse
component centered roughly on DC1842--63. Abell, Corwin and Olowin (1989)
identify Pavo II with S0805. In Figure 10, solid and unfilled circles
identify our best estimate of members and peripheral members of Pavo I,
while filled and unfilled squares do so for Pavo II.

{\it A2634/A2666. } This region has been well studied recently by 
Pinkney \etal (1993) and Scodeggio \etal (1995). We follow the membership
assignment criteria of the latter. A2634 is very well sampled, while A2666
is sufficiently so to allow a good determination of the systemic
velocity and of the velocity dispersion, which indicates a significantly 
less massive character than A2634. Membership assignment to A2634 is 
extended to 2.5$^\circ$ from the center of A2634; within that distance,
however, a foreground subgroup in the A2634 is encountered (A2634--F in
Scodeggio \etal (1995) at $V_{cmb} = 7182$ \kms; the six galaxies associated
with that group in Table 2 are identified by a membership code of ``7''
and by unfilled squares in Figure 12. Galaxies associated with A2666 are
identified by a membership code ``6'' in table 2. The membership assignment
of 476--075 to A2634 is rather marginal; it is just as likely to be a
galaxy in the cluster foreground, projected 0.6$^\circ$ from its
center. An expanded TF sample and distance analysis for this region is
discussed in Scodeggio \etal (1996).

\section {Summary}

We have presented a homogeneous set of data on \ntot spiral galaxies
localized in the fields of 24 groups or clusters of galaxies, which
can be used for the determination of redshift independent distances
using the TF technique. The clusters have systemic velocities between 
about 10$^3$ and 10$^4$ \kms, and range in structural properties from
systems that are relatively evolved, such as Coma, to loose aggregates
such as the nearby groups Ursa Major and MDL59. 

The data include new observations obtained by us and observations
by others, available in the public domain. Photometric properties
are obtained from CCD images in the I band. Spectroscopic parameters
are derived from both single--dish radio observations in the 21 cm
line and from long--slit emission line optical spectra. 

Particular attention has been given to the assignment of cluster
membership. Galaxies have been separated into three main classes:
(a) cluster members, which are assumed to be all at the cluster
redshift ({\bf in} sample); (b) peripheral cluster members, which
are at a redshift very close to that of the cluster but removed
from the cluster center by angular separations larger than about
2 Abell radii; jointly with the objects in class (a) they are
said to be part of the {\bf in+} sample; (c) interlopers, which
include foreground and background objects, and members of peripheral
groups, dynamically separate from the main cluster. Three hundred
and seventy four galaxies comprise the {\bf in} samples and 584 define
the total of the {\bf in+} samples.

We have discussed the characteristics of the errors associated with
the parameters of relevance for the TF analysis, in particular the
error budget for the velocity widths, which contribute an important
source of uncertainty to the TF relation.

These data are used in an accompanying paper (paper VII) to obtain
a TF template relation and to derive the deviations of the individual
clusters from smooth Hubble flow.

\vfill
\acknowledgements

It is a pleasure to thank Dr. M. S. Roberts for carefully reading this 
paper, making suggestions that significantly improved it and for being
kind enough not to comment on its soporific virtues.
The results presented here are based on observations carried out at
the Arecibo Observatory, which is part of the National Astronomy and 
Ionosphere Center (NAIC), at Green Bank, which is part of the National Radio 
Astronomy Observatory (NRAO), at the Kitt Peak National Observatory (KPNO), the 
Cerro Tololo Interamerican Observatory (CTIO), the Palomar Observatory (PO), 
the Observatory of Paris at Nan\c cay and the Michigan--Dartmouth--MIT 
Observatory (MDM). NAIC is operated by Cornell University, NRAO  by  
Associated Universities, inc., KPNO and CTIO by Associated Universities 
for Research in Astronomy, all under cooperative agreements with the National 
Science Foundation. The MDM Observatory was jointly operated by the University 
of Michigan, Dartmouth College and the Massachusetts Institute of Technology 
on Kitt Peak mountain, Arizona. The Hale telescope at the PO is operated by 
the California Institute of Technology under a cooperative agreement with 
Cornell University and the Jet Propulsion Laboratory.  
This research was supported by NSF grants AST94--20505 to RG, AST90-14850 and 
AST90-23450 to MH and AST93--47714 to GW.

%\newpage

%\newpage

%\centerline {\bf Figure Captions}

\begin{figure} % Figure 1
\caption {Stereographic distribution of the 24 clusters listed in Table 1, 
in Cartesian supergalactic coordinates. Labels in the $(X,Y)$ plot identify
clusters; label ``PP'' refers to the three groups in the Perseus--Pisces
supercluster, namely N383, N507 and A262; label ``GA'' refers to the
structures in the Great Attractor region, namely Hydra, Cen30, ESO508, Antlia
and N3557. Dashed circles have radii of 3000 and 6000 \kms, respectively.}
\end{figure}

\begin{figure} % Figure 2
\caption{Sky and velocity distribution of galaxies in the groups
centered around NGC 383 and NGC 507. Filled circles identify galaxies
selected as cluster members ({\bf in} sample), unfilled circles refer 
to galaxies in the cluster periphery (additional members in the 
{\bf in+} sample) and asterisks
identify foreground and background galaxies. Dots give the location
of galaxies with known redshift in the field, but without available
peculiar velocity measurement. Dashed lines in each case 
identify circles of one (1 $R_A$) and two (2 $R_A$) Abell radii; the dotted 
circle in the NGC 383 plot refers to the 1 $R_A$ circle of NGC 507; the 
dotted circle in the NGC 507 plot refers to the 1 $R_A$ circle of NGC 383.}
\end{figure}

\begin{figure} % Figure 3
\caption{Sky and velocity distribution of galaxies in the clusters 
A262 and A400. Filled circles, unfilled circles, asterisks, dots and
dashed circles follow the same convention as in Figure 2.}
\end{figure}

\begin{figure} % Figure 4
\caption{Sky and velocity distribution of galaxies in the clusters 
Fornax and Eridanus. Filled circles, unfilled circles, asterisks, dots and
dashed circles follow the same convention as in Figure 2. Unfilled
triangles in the Fornax sky plot identify Eridanus members; unfilled
triangles in the Eridanus sky plot identify Fornax members.}
\end{figure}

\begin{figure} % Figure 5
\caption{Sky and velocity distribution of galaxies in the clusters 
Cancer and Antlia. Filled circles, unfilled circles, asterisks, dots and
dashed circles follow the same convention as in Figure 2. In the Cancer
plots, unfilled triangles identify objects assigned by Bothun \etal (1983)
to their ``B and C clumps'', while unfilled squares identify galaxies in their
``D clump''.}
\end{figure}

\begin{figure} % Figure 6
\caption{Sky and velocity distribution of galaxies in the Hydra cluster
(A1060) and in the group around NGC 3557. Filled circles, unfilled circles, 
asterisks, dots and dashed circles follow the same convention as in Figure 2.}
\end{figure}

\begin{figure} % Figure 7
\caption{Sky and velocity distribution of galaxies in the clusters 
A1367 and Ursa Major. Filled circles, unfilled circles, asterisks, dots and
dashed circles follow the same convention as in Figure 2. Caustics in the
lower panel of A1367 are plotted for a value of the cosmological density
parameter $\Omega_\circ = 0.3$.}
\end{figure}

\begin{figure} % Figure 8
\caption{Sky and velocity distribution of galaxies in the clusters Centaurus 
(A3526) and Coma (A1656). Filled circles, unfilled circles, asterisks, dots and
dashed circles follow the same convention as in Figure 2. We identify the
Centaurus (A3526) cluster with the foreground component Cen30 (Lucey \etal 
1986).  Unfilled squares in the Centaurus panels identify galaxies that have 
been assigned to the background cluster Cen45. Caustics in the lower panel of 
A1656 are plotted for a value of the cosmological density parameter 
$\Omega_\circ = 0.3$.}
\end{figure}

\begin{figure} % Figure 9
\caption{Sky and velocity distribution of galaxies in the clusters ESO 508 
and A3574 (Klemola 27). Filled circles, unfilled circles, asterisks, dots and
dashed circles follow the same convention as in Figure 2. The cluster 
NW of center of A3574 and SE of center of ESO 508 is A1736, a background 
structure at $z \sim 0.04$. Unfilled
squares in the plots of A3574 identify three galaxies in the group S753.}
\end{figure}

\begin{figure} % Figure 10
\caption{Sky and velocity distribution of galaxies in the clusters 
A2197, A2199, Pavo  and Pavo II (S0805). In the A2197/99 plots, filled circles
and triangles identify members of A2199 and A2197, respectively. Unfilled
circles identify peripheral members of the binary system. Pavo II
members are identified by filled squares, while Pavo members are displayed
as filled circles; unfilled squares and circles identify peripheral members
of, respectively, Pavo II and Pavo. The systemic velocities of A2199 and Pavo II
are indicated by horizontal dashed lines, respectively in the two lower panels 
of the figure.}
\end{figure}

\begin{figure} % Figure 11
\caption{Sky and velocity distribution of galaxies in the clusters 
MDL59 and Pegasus. Filled circles, unfilled circles, asterisks, dots and
dashed circles follow the same convention as in Figure 2.}
\end{figure}

\begin{figure} % Figure 12
\caption{Sky and velocity distribution of galaxies in the clusters 
A2634 and A2666. Filled circles, unfilled circles, asterisks, dots and
dashed circles follow the same convention as in Figure 2. The dotted circle
in the A2634 sky plot is the 1 $R_A$ circle 
of A2666; likewise, the dotted circle in the A2666 plot
is at 1 $R_A$ for A2634. Unfilled squares in the
plots of both clusters refer to objects in a group in the foreground of 
A2634 near 7,000 \kms; this group is closer to A2634, in projection on the
sky, and closer in redshift to A2666. Foreground and background objects are 
only plotted in the A2634 figures.}
\end{figure}

\begin{figure} % Figure 13
\caption{Distribution of I band ellipticity errors, plotted as a function of 
the measured ellipticities, for objects with CCD photometry measured
by us. Galaxies pertain to the cluster sample presented in this paper and to 
a field sample of late--type spirals. The solid line represents eqn. (4) in 
the text, an estimate of the median value of the error, which can be applied
to data sets for which ellipticity errors are not reported.}
\end{figure}

\begin{figure} % Figure 14
\caption{Fractional error on the velocity width of 2000 cluster and
field spirals, plotted versus the corrected velocity width.}
\end{figure}

\begin{figure} % Figure 15
\caption{Magnitude residuals computed with respect to a TF template relation,
obtained in paper VII, computed before the internal extinction carrection
is applied to the data. Data points refer to running averages for the
all sky spiral sample discussed in Papers I and II (filled circles) and 
for the cluster sample ({\bf In+}) introduced in this paper (unfilled squares).
The mean residuals are plotted versus the mean log of the axial ratio 
(corrected for seeing), an indication of the disks' 
inclination to the line of sight. Solid lines are {\it not} fits to the
data; they represent instead the internal extinction correction law, eqn. (12),
resulting from averaging the value of $\gamma$ for all the galaxies in 
subsamples (c) through (e); see text for details on bilinear model for (a) 
and (b). $\gamma$ gives the slope of the extinction line. Panels (a) to (e)
refer to non--overlapping bins in $\log W_c$, i.e. galaxies with $\log W_c > 2.60$
(a), 2.5 to 2.6 (b), 2.4 to 2.5 (c), 2.3 to 2.4 (d) and $\log W_c < 2.3$ (e). }
\end{figure}

\begin{figure} % Figure 16
\caption{Apparent, raw I band magnitude distribution of all galaxies in 
the {\bf in+} cluster samples, including cluster members and peripheral
objects.}
\end{figure}

\begin{figure} % Figure 17
\caption{Distribution of isophotal angular radii of all galaxies in the 
{\bf in+} cluster sample, including cluster members and peripheral objects.}
\end{figure}

\begin{figure} % Figure 18
\caption{Absolute I band magnitude distribution of all galaxies in the 
{\bf in+} cluster sample, including cluster members and peripheral 
objects. The solid line superimposed on the histogram represents the spiral 
galaxy luminosity function from which the observed sample is assumed to be
extracted, parametrized by a Schechter formula with $M^*_I + 5\log h=-21.6$ 
and  $\alpha = -0.50$.}
\end{figure}
\vfill
\end{document}